\documentclass[final,5p,times,twocolumn]{elsarticle}
\usepackage{graphicx,amssymb,dcolumn,bm,amsmath,natbib}
\usepackage{indentfirst}
\usepackage{epsfig}
\usepackage{epstopdf}
\usepackage{exscale}
\usepackage{caption2}
\usepackage{relsize}
\usepackage{xcolor} 

\journal{Chaos, Solitons \& Fractals}
\biboptions{sort&compress}
\biboptions{}

\begin{document}

\begin{frontmatter}

\title{Dynamics of quasi-one-dimensional quantum droplets in Bose-Bose mixtures}

\author{Sherzod R. Otajonov \corref{cor1}}
\address{Physical-Technical Institute of the Uzbek Academy of Sciences,\\
Chingiz Aytmatov str. 2-B, Tashkent, 100084, Uzbekistan}

\address{National University of Uzbekistan, Department of Theoretical Physics,\\ 100174, Tashkent, Uzbekistan}

\cortext[cor1]{Corresponding author.}
\ead{srotajonov@gmail.com}

\author{Bakhram A. Umarov}
\address{Physical-Technical Institute of the Uzbek Academy of Sciences,\\
Chingiz Aytmatov str. 2-B, Tashkent, 100084, Uzbekistan}

\author{Fatkhulla Kh. Abdullaev}
\address{Physical-Technical Institute of the Uzbek Academy of Sciences,\\
Chingiz Aytmatov str. 2-B, Tashkent, 100084, Uzbekistan}

\address{Institute of Theoretical Physics, National University of Uzbekistan,\\ 100174, Tashkent, Uzbekistan}

\date{\today}




\begin{abstract}
 The properties of quasi-one-dimensional quantum droplets of Bose-Einstein condensates are investigated analytically and numerically, taking into account the contribution of quantum fluctuations. Through the development of a variational approach employing the super-Gaussian function, we identify stationary parameters for the quantum droplets. The frequency of breathing mode oscillations in these quantum droplets is estimated. Moreover, the study reveals that periodic modulation in time of the atomic scattering length induces resonance oscillations in quantum droplet parameters or the emission of linear waves, contingent on the amplitude of the external modulation. A similar analysis is conducted for the Lee-Huang-Yang fluid, confined in a parabolic potential. Theoretical predictions are corroborated through direct numerical simulations of the governing extended Gross-Pitaevskii equation. Additionally, we study the collision dynamics of quasi-one-dimensional quantum droplets.
\end{abstract}


\end{frontmatter}

\section{Introduction}	
\label{sec:intro}

One of the breakthrough achievements in condensed matter physics at the end of last century was the experimental realization of Bose-Einstein condensates (BECs) in dilute atomic gases~\cite{Cornell, Ketterle, Pethick2008, Pitaevskii2016}. It has been shown that mean-field theory, based on the Gross-Pitaevskii equation (GPE), can be successfully applied to explain the existence, stability, and dynamics of BEC~\cite{Pethick2008, Pitaevskii2016, Abdullaev2005}.
Recently, the investigation of beyond mean-field effects has become an actual topic in the physics of atomic Bose-Einstein condensates~\cite{Malomed2020}. One particularly interesting consequence of this research is the discovery of quantum droplets (QDs), which have attracted the attention of many researchers. Quantum droplets are self-bound states of BEC, existing in three-dimensional free space due to the balancing mechanism coming from the competition between the attractive mean-field energy and repulsive beyond mean-field energy, Lee-Huang-Yang (LHY) corrections~\cite{LHY}.
Formation of the quantum droplet was first theoretically predicted by Petrov~\cite{Petrov2015} for two-component Bose gas. It is known that in mean-field theory BEC of attracting atoms is unstable against collapse in two and three dimensions~\cite{Pethick2008, Pitaevskii2016}. Stable BEC exists in one-dimension~\cite{Hulet2002}, when attracting forces between atoms are compensated by dispersion because of quantum pressure mixtures. 
One of the main motivations for the research of quantum droplets came from the search for stable BEC states in two and three spatial dimensions~\cite{Malomed2022}. The effect of the quantum corrections on the ground state energy of the one component Bose gas, described by the LHY term, is negligibly small compared to the contribution of mean field terms in the governing Gross-Pitaevskii equation. However, as proposed by Petrov~\cite{Petrov2015}, in symmetric binary BECs, the interspecies and intraspecies interactions can be manipulated in such a way that the final single modified GPE will contain both the attractive mean-field term and the repulsive LHY term.  Importantly, the strength of the attractive term can be controlled independently, allowing the formation of dilute liquid droplet states stabilized by quantum fluctuations.
The modified three-dimensional GPE equation~\cite{Petrov2016}, with repulsive quartic nonlinearity responsible for LHY corrections, was derived and successfully applied to describe stable quantum droplets in BEC. The first experimental realization of QD took place for single component Bose-Einstein condensates (BEC) with dipolar atomic interactions ~\cite{ferrier-barbut2016,kadau2016}, and later in binary Bose gas mixtures~\cite{Cabrera2018, Cheyney2018, Semeghini2018}. Further experimental studies with quantum fluctuations include QD in a heteronuclear bosonic mixture ~\cite{DErrico2019}, and collision of the droplets~\cite{Ferioli2019}.
The existence of quantum droplets in lower dimensions also was elaborated. In particular, it was shown that in one dimension the tight external potential strongly changes the spectrum of zero-point excitations. It leads to the modification of the LHY term from quartic to quadratic and changing of its sign~\cite{Zin2018, Ilg2018, Lavoine2021}. Under those circumstances, to provide the existence of quantum droplets, two-body mean-field interaction should be repulsive. The existence, stability and dynamics of quantum droplets for 1D GPE in this case were discussed in                   
~\cite{Otajonov2019, Astrakharchik2018, Kartashov2022, Kevrekidis2023, Debnath2023-2}. 
 On the other side, if the external potential is loose, the LHY corrections are the same as obtained in the 3D case ~\cite{Zin2018}. The quasi-1D GPE and quantum droplets in this case have been discussed in the papers~\cite{Debnath2021, Debnath2022}. The surface tension of QD in 1D and quasi-1D ultra-cold atomic Bose gases is analyzed in~\cite{Khan2024} and the paper~\cite{Debnath2023-1} is devoted to the interaction of quasi-1D QD with impurities.

The objective of our work is to study further analytically and numerically the quasi-1D quantum droplets in the two-component BEC. The paper is organized as follows. In the first section of the paper, the quasi-1D GPE is introduced and the physical meaning of parameters is explained. Then we apply variational approximation (VA) with super-Gaussian ansatz to find an approximate solution for quantum droplets and thoroughly discuss the properties of these solutions. The dynamical equations for the parameters of quantum droplets are derived. In the next two sections, we consider the quantum droplets under periodic modulations of the atomic scattering length and the case of LHY fluid which is formed, when the two-body attraction is reduced to zero and the repulsive term in GPE caused by quantum fluctuations balanced by external potential. In the last section, the collisions of quantum droplets are studied via numerical simulation of quasi-1D GPE.

\section{The model and Variational approach}
\label{sec:model}

Let us consider a three-dimensional, two-component Bose-Einstein condensate influenced by quantum fluctuations. Here, both the mass and the number of atoms in each component are set to be identical. Under these conditions, the system can be described by a single Gross-Pitaevskii equation~\cite{Petrov2015}:
\begin{eqnarray}
& i \hbar \cfrac{\partial \Phi}{\partial T} + \cfrac{\hbar^2}{2 m_0} \nabla^2 \Phi -U(x,y,z) \Phi + \gamma_{\mathrm{3D}} |\Phi|^2 \Phi 
\nonumber \\
& - \delta_{\mathrm{3D}} |\Phi|^3 \Phi=0\,.
\label{dimGPE3D}
\end{eqnarray}
In this context, $\Phi=\Phi(x,y,z,t)$ symbolizes the condensate wave function, with $|\Phi|^2$ indicating the density of the condensate. Parameter $\gamma_{\mathrm{3D}}=4 \pi \hbar^2 \delta a /m_0$ and $\delta_{\mathrm{3D}}=256 \sqrt{2 \pi} \hbar^2 a^{5/2} / 3 m_0$ represent the strengths of the two-body interactions and quantum fluctuations, respectively. The Laplacian operator, $\nabla^2$, represents the spatial derivatives across all three dimensions. Here, $T$ denotes time, and $m_0$ represents the atomic mass. The parameter $\delta a=a+a_{12}$ describes the residual mean-field interaction, where $a_{11}=a_{22}=a$ and $a_{12}$ are the intra- and inter-species scattering lengths, respectively.

The external potential, $U(x,y,z)=m_0 \omega_{\perp}^2 r_{\perp}^2 / 2+m_0 \omega_{x}^2 x^2 / 2$, incorporates the longitudinal trap frequency $\omega_x$, the transverse trap frequency $\omega_{\perp}$ (where $\omega_y = \omega_z$), and $r_{\perp}^2=y^2+z^2$.
We focus on the situation where $\omega_x / \omega_{\perp} \ll 1$, which indicates a condensate with a cigar-shaped geometry. In this particular arrangement, the condensate is subject to tight confinement along two axes, effectively creating a quasi-one-dimensional system~\cite{Barenghi2016, Pitaevskii2016}. Consequently, we can employ a factorization technique as introduced in Ref.~{\cite{Michinel1997}}, which ultimately guides us towards the quasi-1D Gross-Pitaevskii equation: 
\begin{eqnarray}
& i \hbar \cfrac{\partial \phi}{\partial T} + \cfrac{\hbar^2}{2 m_0} \cfrac{\partial^2 \phi}{\partial x^2} -U(x) \phi + \gamma_{\mathrm{1D}} |\phi|^2 \phi 
\nonumber \\
& - \delta_{\mathrm{1D}} |\phi|^3 \phi=0\,,
\label{dimGPE1D}
\end{eqnarray}
where $\gamma_{\mathrm{1D}}=\gamma_{\mathrm{3D}}/2 \pi l_{\perp}^2$ and $\delta_{\mathrm{1D}}=2 \delta_{\mathrm{3D}} /5 \pi^{3/2} l_{\perp}^3$, where $l_{\perp}=(\hbar/m_0 \omega_{\perp})^{1/2}$ is harmonic oscillator length.

Through the application of the rescaling transformations $t = T/t_S$, $x = x/x_s$, and $\Psi = \phi/\Psi_s$, one can express Eq.(\ref{dimGPE1D})  in a dimensionless form: 
\begin{equation}
  i \Psi_t + \cfrac{1}{2} \Psi_{xx} - V(x) \Psi 
 + \gamma |\Psi|^2 \Psi - \delta |\Psi|^3 \Psi = 0,
\label{quasi_1D_gpe} 
\end{equation}
where $V(x) = \alpha x^2$ represents the external potential after rescaling with $\alpha =m_0 \omega_x^2 x_s^2 t_s / 2 \hbar$, the scale parameters are defined as follows:
$$t_s=\cfrac{2^{16}\, m_0 \gamma^3 a^5}{225\, \pi^2 \hbar \, \delta^2 \delta a^3 }\,, \quad
x_s=\cfrac{256}{15 \pi} \left( \cfrac{\gamma^3 a^5}{\delta^2 \delta a^3} \right)^{1/2}\,,$$  
$$\psi_s=\cfrac{15 \pi \delta a \, l_{\perp} \delta }{256 \sqrt{2}\, a^{5/2} \gamma } \, .$$
It is possible to normalize Eq.~(\ref{quasi_1D_gpe}) so that the coefficients satisfy $|\gamma| = |\delta| = 1$.
Nonetheless, we maintain these notations in the dimensionless equation to facilitate the analysis of various scenarios, encompassing cases where either $\gamma$ or $\delta$ equals zero. 

To examine the dynamics of the quasi-1D BEC influenced by quantum fluctuations, we find it advantageous to employ the variational approach reported in Refs.~\cite{Otajonov2019, Otajonov2020, Otajonov2022_1, Lavoine2021}.
The Lagrangian density corresponding to Eq.~(\ref{quasi_1D_gpe}) is formulated as
\begin{eqnarray}
   & \mathcal{L} = \cfrac{i}{2} (\Psi_t^{*} \Psi - \Psi^{*} \Psi_t) + \cfrac{1}{2} |\Psi_x|^2 + V(x) |\Psi|^2 
\nonumber \\
 & -\cfrac{\gamma}{2}  |\Psi|^4 + \cfrac{2 \delta}{5} |\Psi|^5.
\label{lagr} 
\end{eqnarray}

In our analysis, we use a super-Gaussian trial function to approximate the density distribution: 
\begin{equation}
  \Psi(x,t) = A \exp \left[ - \cfrac{1}{2} \left( \cfrac{x}{w}\right)^{2m} +
     i b x^2 + i \varphi  \right] .
\label{sgauss}
\end{equation}
where $A(t)$ is the amplitude, $w(t)$ is the width, $b(t)$ is the chirp, and $\varphi(t)$ is the linear phase of the condensate. These parameters are time-dependent, while the super-Gaussian parameter $m$ is found from the stationary solution, see Eq.~(\ref{fb_fm}). The application of the super-Gaussian trial function to the Gross-Pitaevskii type equation has been also previously employed in several studies~\cite{Karlsson1992, Baizakov2011}. In this work, we primarily investigate the existence and stability of standing solutions. Including the centre-of-mass coordinate and velocity in the VA is feasible and beneficial for exploring soliton dynamics in trapping potentials, collective oscillations, and similar phenomena (see for example ~\cite{Debnath2023-2, Debnath2023-1}). However, this is the reason we do not incorporate the centre-of-mass coordinate in Eq.~(\ref{sgauss}).

The norm is a conserved quantity directly proportional to the number of atoms in the BEC. Its value can be found in the integral of the module square of the condensate wave function~\cite{Barenghi2016}. 
\begin{equation}
   N \equiv \int_{-\infty}^{\infty} |\Psi|^2 dx = 2\Gamma(M+1) A^2 w,
\label{norm}
\end{equation}
where $M = 1/(2m)$ is reduced super-Gaussian indeces, and $\Gamma(z)$ is the Gamma function. The parameter $N = N_{\mathrm{real}} / N_s$ is the dimensionless parameter related to the real number ($N_{\mathrm{real}}$) of atoms in a BEC cloud, where $N_s =\psi_s ^2 x_s$ is the scale parameter. All theoretical and numerical results pertain to the dimensionless equation. However, at the end of Sec. 4, we also present parameters in physical units. 

The Lagrangian function is found from the integral: $L = \int_{-\infty}^{\infty} \!\!\! \mathcal{L}\, dx $, 
\begin{eqnarray}
 & \cfrac{L}{N}  = \varphi_t + \cfrac{\Gamma(3M)}{\Gamma(M)}\, w^2 (2 b^2 + b_t +\alpha )
  + \cfrac{\Gamma(2-M)}{8 M^2\, \Gamma(M)\, w^2}
\nonumber \\
 &  - \cfrac{2^{-M-2} \gamma  N}{\Gamma(M+1)\, w} + \left(\cfrac{2}{5}\right)^{M+1} \delta \left( \cfrac{N}{2 \Gamma(M+1) w} \right)^{3/2}.
\label{avlagr}
\end{eqnarray}
In obtaining Eq.~(\ref{avlagr}), we employ the expression (\ref{norm}) for the norm to eliminate the amplitude. 
The gamma function $\Gamma (2-M)$ incorporates an argument of $2-M$, which must be positive. This stipulation imposes an additional constraint on $m$, specifically $m>0.25$. The Euler-Lagrange equations for $\varphi$ give the $dN/dt=0$, which refers to the conservation of the number of atoms. The equations of motion for $w$ and $b$ yields equations governing the time derivatives $b_t$ and $w_t$, respectively: 
\begin{eqnarray}
   && b_t = -\alpha -2 b^2 + \cfrac{1}{8 \Gamma(3 M+1)} \left[ \cfrac{3 \Gamma(2-M)}{M w^4}  \right. 
\nonumber \\
     && \left. -\cfrac{3 \gamma N}{2^{M} w^3} +
     \cfrac{9 \delta N}{w^3} \left( \cfrac{2}{5} \right)^{M+1} \left( \cfrac{N}{2 \Gamma(M+1) w} \right)^{1/2}
     \right] \equiv f_b,
\nonumber \\
   && w_t = 2 b w \equiv f_w \,.
\label{dyneq}
\end{eqnarray}
We employ the given equation to determine the stationary profile of the QDs. In the stationary scenario, the quantum droplet's profile remains constant over time, implying that the time derivative should be equal to zero. The latter of Eq.~(\ref{dyneq}) yields $b=0$. Consequently, the stationary parameters can be obtained from the following equations:
\begin{eqnarray}
   f_b(w, m, N) &=& 0,
\nonumber \\
   f_m(w, m, N) &\equiv& \left. \cfrac{\partial L}{ \partial m} \right|_{b=0}= 0.
\label{fb_fm}
\end{eqnarray}
The former of Eq.~(\ref{fb_fm}) is a quadratic equation on $w$. Only the one root is real and positive. By solving the equations (\ref{fb_fm}) for given parameters $(\gamma, \delta, N)$ the stationary values of QD width $w_s$ and super-Gaussian indices $m_s$ can be obtained. Subsequently, the stationary amplitude of QDs $A_s$ is obtained from Eq.~(\ref{norm}). We find that the dynamics of droplets are well described by assuming $m = m_s$.

The set of equations~(\ref{dyneq}) can be simplified to a single equation concerning the width of the QD, given by 
$$w_{tt} = 4b^2w + 2b_t w\,.$$
Then, the effective potential is found from $w_{tt}=-\partial U(w) / \partial w$ relation, which resembles the equation of motion for a unit mass particle in an anharmonic potential.
\begin{eqnarray}
   U(w) =
      \alpha w^2 + \cfrac{1}{\Gamma(3 M)} \left[
     \cfrac{\Gamma(2-M)}{8 M^2 w^2} - \cfrac{\gamma N}{ 2^{M+2} M w} 
     \right.
\nonumber \\
      \left.
     +\left( \cfrac{2}{5} \right)^{M+1} \cfrac{\delta N}{M w} \left( \cfrac{ N}{8 \Gamma(M+1) w} \right)^{1/2}
     \right].
\label{pot}
\end{eqnarray}

\begin{figure}[htbp]
  \centerline{ \includegraphics[width=4.55cm]{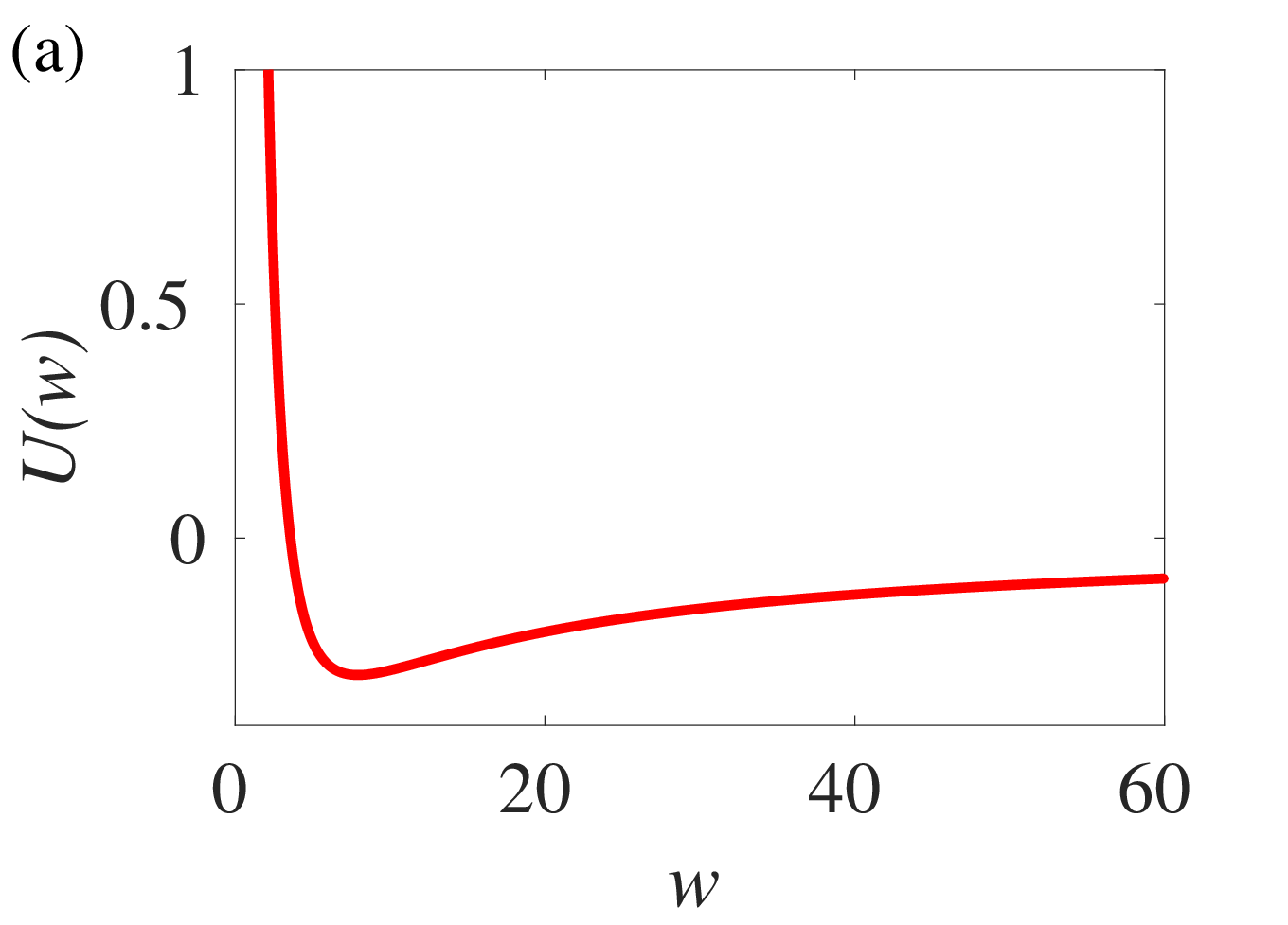} \includegraphics[width=4.55cm]{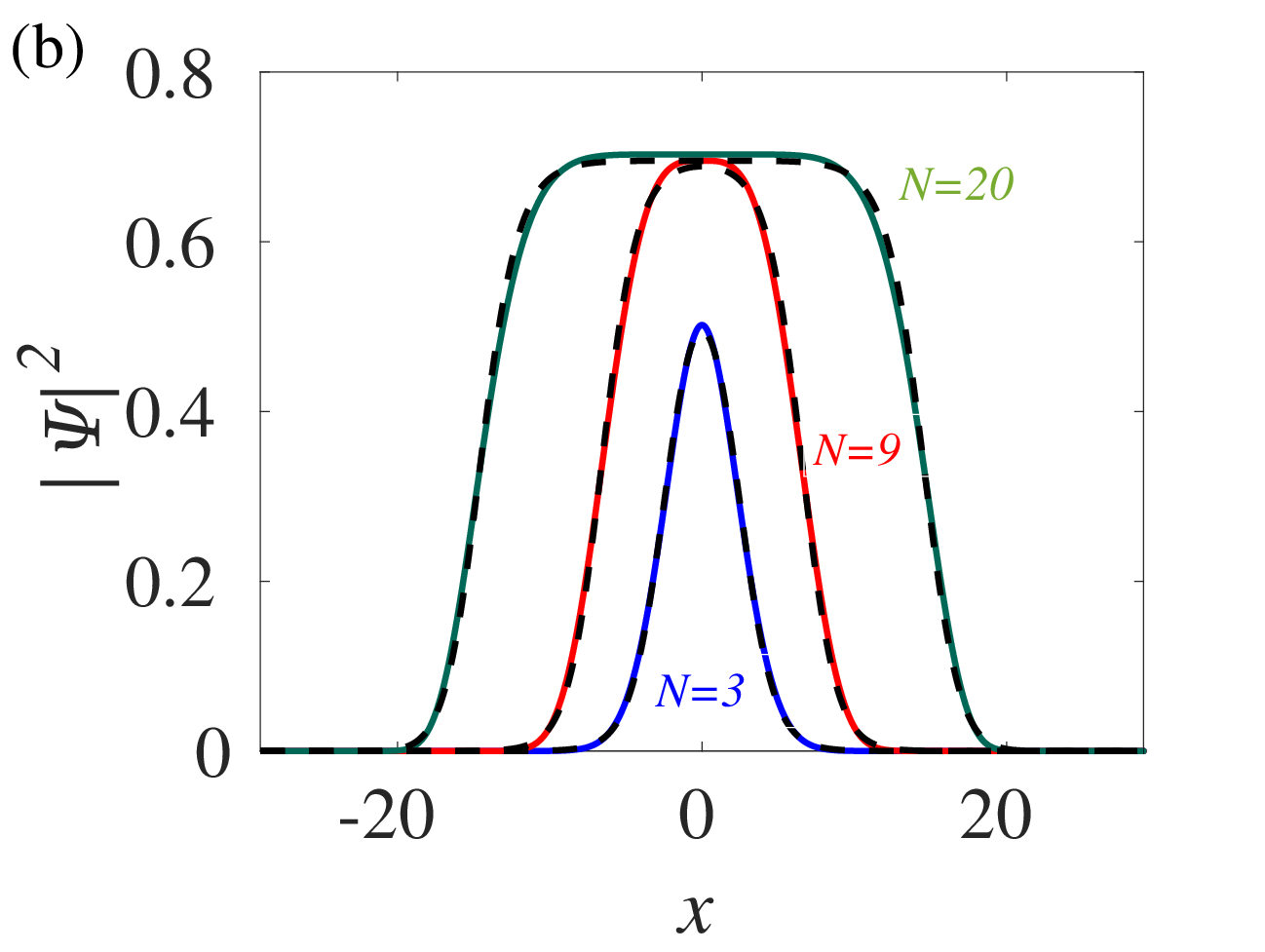}}
\caption{(a) The characteristic form of the effective potential $U(w)$, corresponding to $N=3$. 
(b) The profiles of stationary QDs for different values of $N$, straight lines are determined through the VA, while dashed lines are obtained from imaginary time simulations Eq.~(\ref{quasi_1D_gpe}). Other parameters ($\alpha, \gamma, \delta) =(0,1,1)$. }
\label{fig-1ab}
\end{figure}

The figure~\ref{fig-1ab}(a) illustrates the typical shapes of the effective potentials. The minimum point of the potential curve, or where $f_b(w, m, N)$ equals zero, corresponds to the stationary width $w_s$ of the QDs. 
Figure~\ref{fig-1ab}(b) compares the stationary profiles predicted by VA and the solutions obtained numerically using the imaginary time method.

   The chemical potential $\mu$ for the stationary QDs can be determined as:
\begin{eqnarray}
   \mu &= \cfrac{\partial E_{s} }{\partial N}
     = \cfrac{2 \alpha \Gamma(3 M)}{\Gamma(M)} w^2_s - \cfrac{3 \gamma N}{2^{M+3} M \Gamma(M)\, w_s}
\nonumber \\     
 & + \cfrac{ 7 \delta}{10} \left( \cfrac{2}{5} \right)^{M} \left( \cfrac{N}{2 w_s \Gamma(M+1)} \right)^{3/2},
\label{mu}
\end{eqnarray}
where
\begin{equation}
E_s=\int  \limits_{\ -\infty}^{\ \ \infty} dx \left( \cfrac{1}{2} |\Psi_x|^2 + V(x) |\Psi|^2 - \cfrac{\gamma}{2}  |\Psi|^4 + \cfrac{2 \delta}{5} |\Psi|^5 \right).
\label{statEnergy}
\end{equation}
is the stationary energy of the QD.

\begin{figure}[htbp]
  \centerline{ \includegraphics[width=4.55cm]{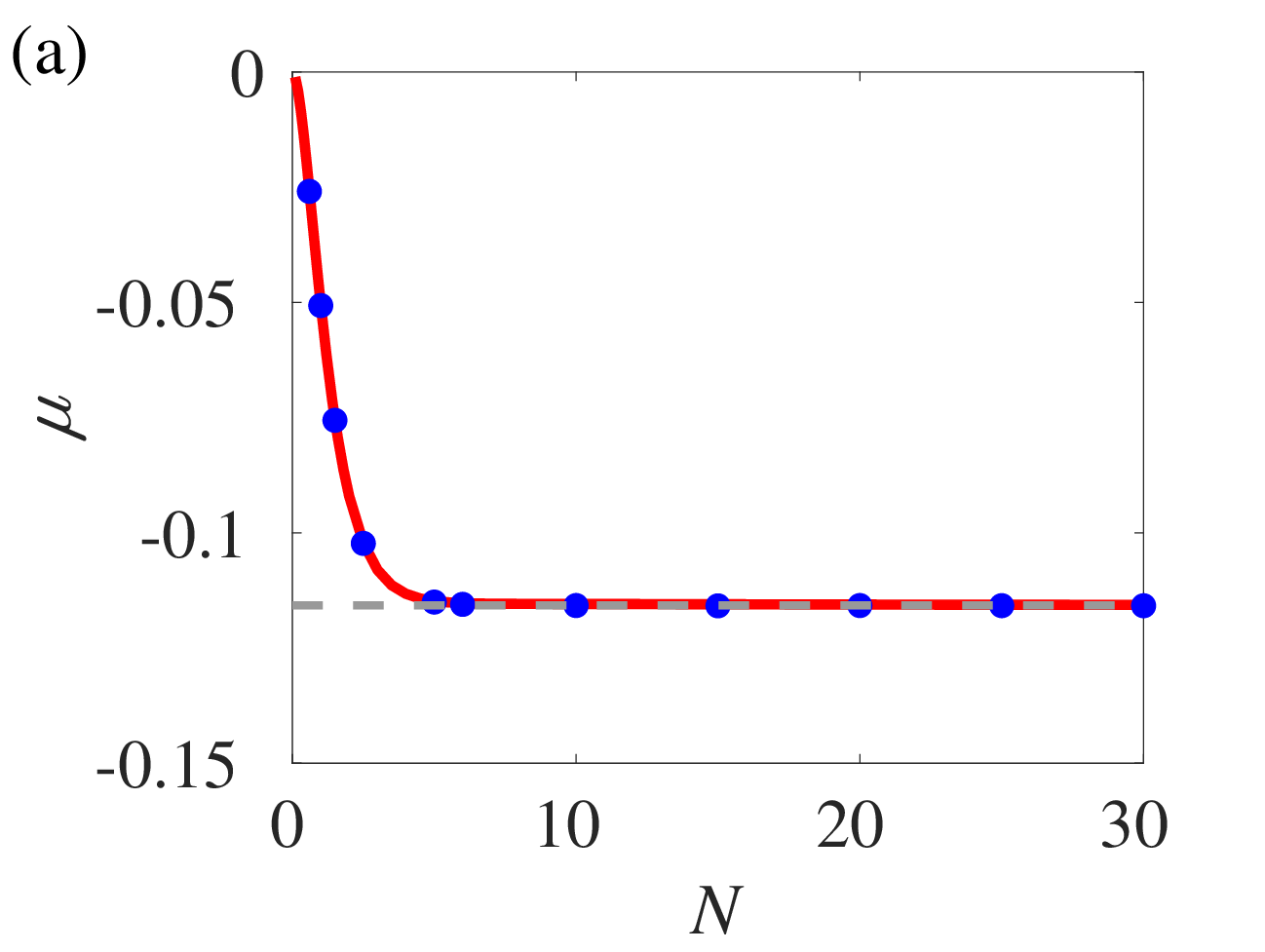} \includegraphics[width=4.55cm]{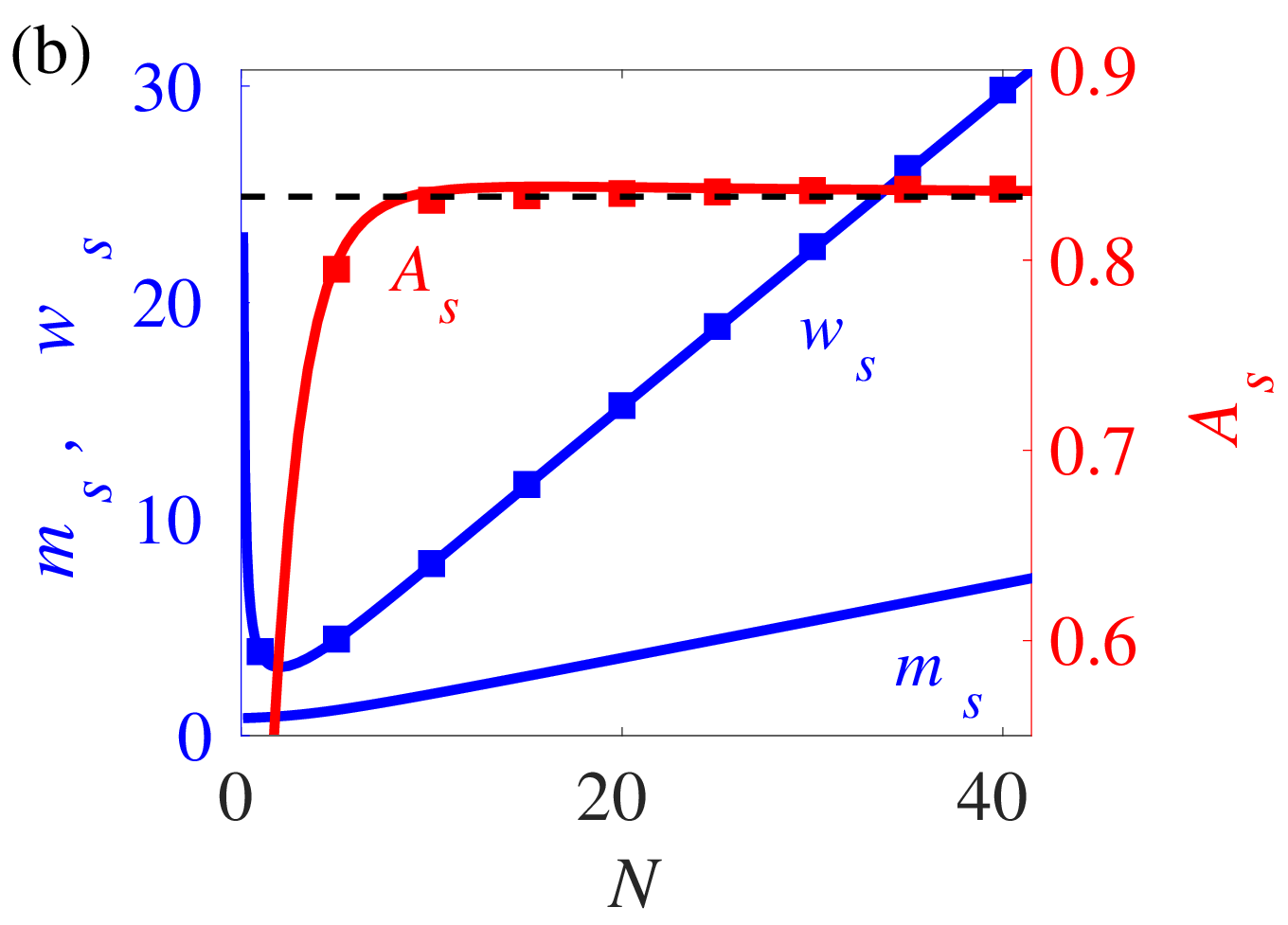}}
\caption{(a) The dependence between the chemical potential $\mu$ and the particle number $N$. The dashed line represents the Thomas-Fermi limit of a chemical potential $\mu_{\mathrm{TF}}$. (b) The stationary QD parameters for different values of $N$. The dashed line represents the Thomas-Fermi limit of QD amplitude $A_{\mathrm{TF}}$ (right axis). In both panels, the lines are determined through the VA, whereas the points correspond to imaginary time simulations of Eq.~(\ref{quasi_1D_gpe}). Other parameters ($\alpha, \gamma, \delta) =(0,1,1)$.}
\label{fig-2ab}
\end{figure}

To investigate the stability of stationary solutions, we first checked the Vakhitov-Kolokolov stability criterion, which is often used to analyze the stability of solutions in nonlinear systems. It involves examining the sign of the chemical potential derivative concerning the norm. According to this criterion, the system is considered stable when $d\mu/dN < 0$. This implies that an augmentation in the number of atoms corresponds to a reduction in the chemical potential, indicating a stable behaviour for the stationary solutions. One can see from the $\mu(N)$ dependence in Fig.~\ref{fig-2ab}(a) that, an augmentation in the number of atoms corresponds to a reduction in the chemical potential, satisfies the Vakhitov-Kolokolov criteria and indicates a stable behaviour for the stationary quantum droplets. In general, the Vakhitov-Kolokolov criterion serves as a necessary but not sufficient condition for the stability of localized solutions. Hence, the theoretical results were systematically tested through direct numerical simulations of the governing Eq.~(\ref{quasi_1D_gpe}). The results robustly indicate that all quantum droplets are stable against small perturbations.
%

	Figure~\ref{fig-2ab}(b) presented a comparison between parameters determined through the VA (represented by lines) and those found from direct numerical simulations of Eq.~(\ref{quasi_1D_gpe}) (depicted as points). 
Figures~\ref{fig-1ab}(b) and \ref{fig-2ab} reveal that the VA employing the super-Gaussian function yields good results for the main properties of QDs.  The proposed trial function (\ref{sgauss}) includes an additional parameter, $m$, compared to the standard Gaussian function (where $m=1$ in Eq.~(\ref{sgauss})). This additional super-Gaussian parameter allows this function to describe both bell-shaped and flat-top characteristics, which are typical of QDs. The standard Gaussian function is limited to describing bell-shaped states with a very small number of atoms and cannot account for flat-top states.

When the number of atoms is large, the density of the QDs attains a uniform saturation density with low compressibility, resulting in bulk energy characteristic of flat-top QDs. These values can be approximated using the Thomas-Fermi (TF) limit, as shown in Ref.~\cite{Li2018}. In the case where there is no external potential, $V(x)=0$, we apply the TF approximation. The energy density of the flat region, which has the peak density $n_p \equiv |\Psi_p|^2$ in the QD profile, is given by $\varepsilon (n_p)=-\gamma n_p^2 / 2 + 2 \delta n_p^{5/2} /5$, as described by Eq.~(\ref{statEnergy}). The minimization of the bulk energy, $d E_{\mathrm{bulk}}/d n_{\mathrm{max}}=0$ yields $n_{\mathrm{TF}}^{1/2}=5\, \gamma /6\, \delta$. This corresponds to an amplitude $A_{\mathrm{TF}}=n_{\mathrm{TF}}^{1/2}$ and a chemical potential $\mu_{\mathrm{TF}} =d \varepsilon(n_{\mathrm{max}})/d n_{max}= -\gamma\, n_{\mathrm{TF}} + \delta \,n_{\mathrm{TF}}^{3/2}$. 
Figure \ref{fig-2ab} clearly shows that for fixed values of $\gamma=\delta=1$, the droplet amplitude and chemical potential tend to approach the Thomas-Fermi limit for a large number of atoms, with $A_{\mathrm{TF}}=5/6$ and $\mu_{\mathrm{TF}}=-(5/6)^2 /6$, respectively. It is important to note that these TF limits can also be determined by calculating the asymptotic behaviour of the equations derived from the VA. 
An increase in the super-Gaussian parameter $m_s$ with larger $N$ suggests that the density profile of a QD transitions towards a flat-top shape. The functional dependence of $m_s$ on $N$ can be approximated by $m_s(N) = 0.10214(N + 5.0424)^{1.1051}$. This relationship, capturing the behaviour of $m_s(N)$, has been accurately described for $N$ values in the range $0<N \leq 100$, employing a power law fit of the form $m_s(N) = a_1(N + a_2)^k$. The droplet width exhibits nearly linear growth with increasing N. This property of QDs displays an incompressibility reminiscent of ordinary liquids. 

\begin{figure}[htbp]
  \centerline{ \includegraphics[width=4.55cm]{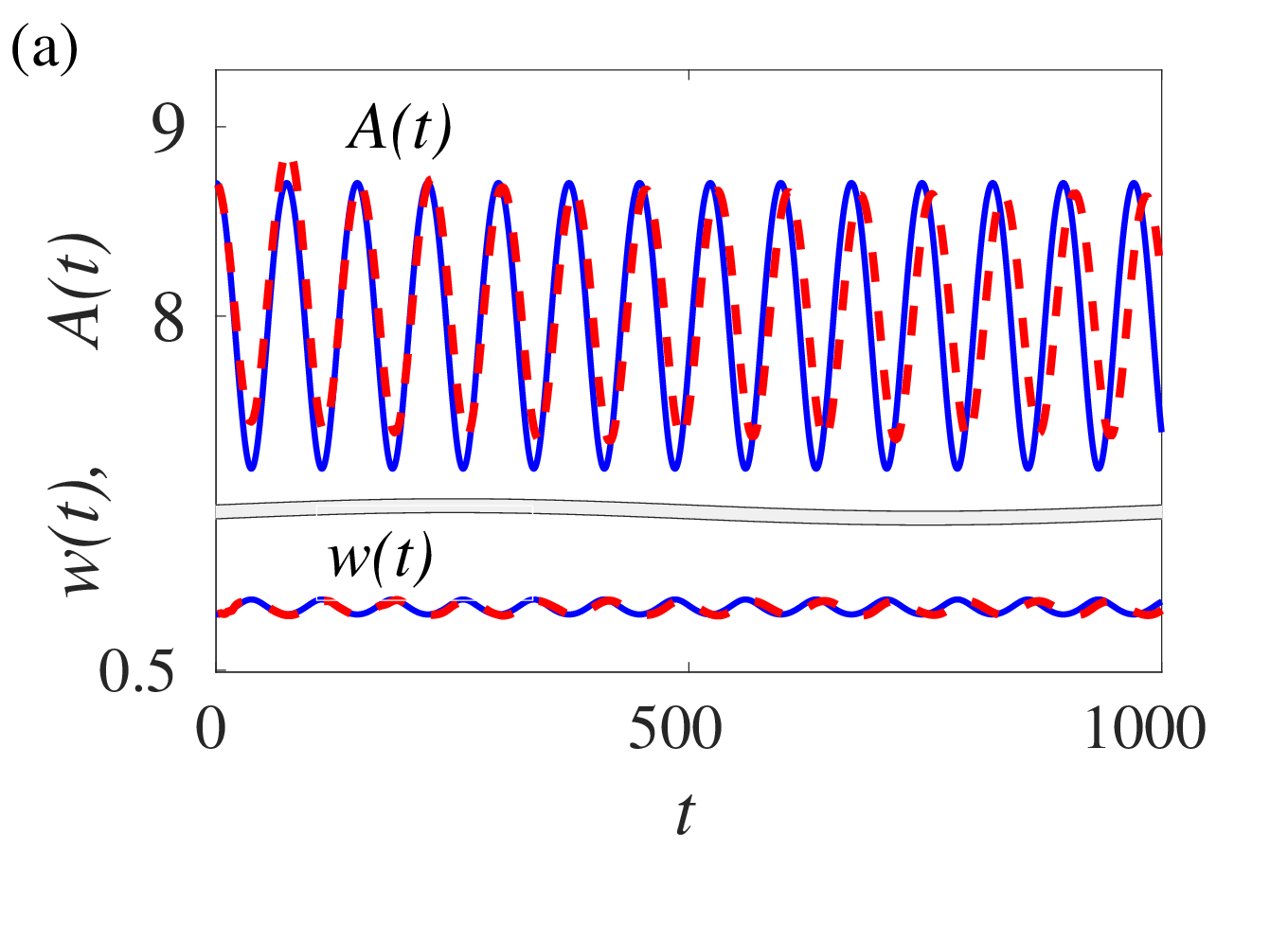} \includegraphics[width=4.55cm]{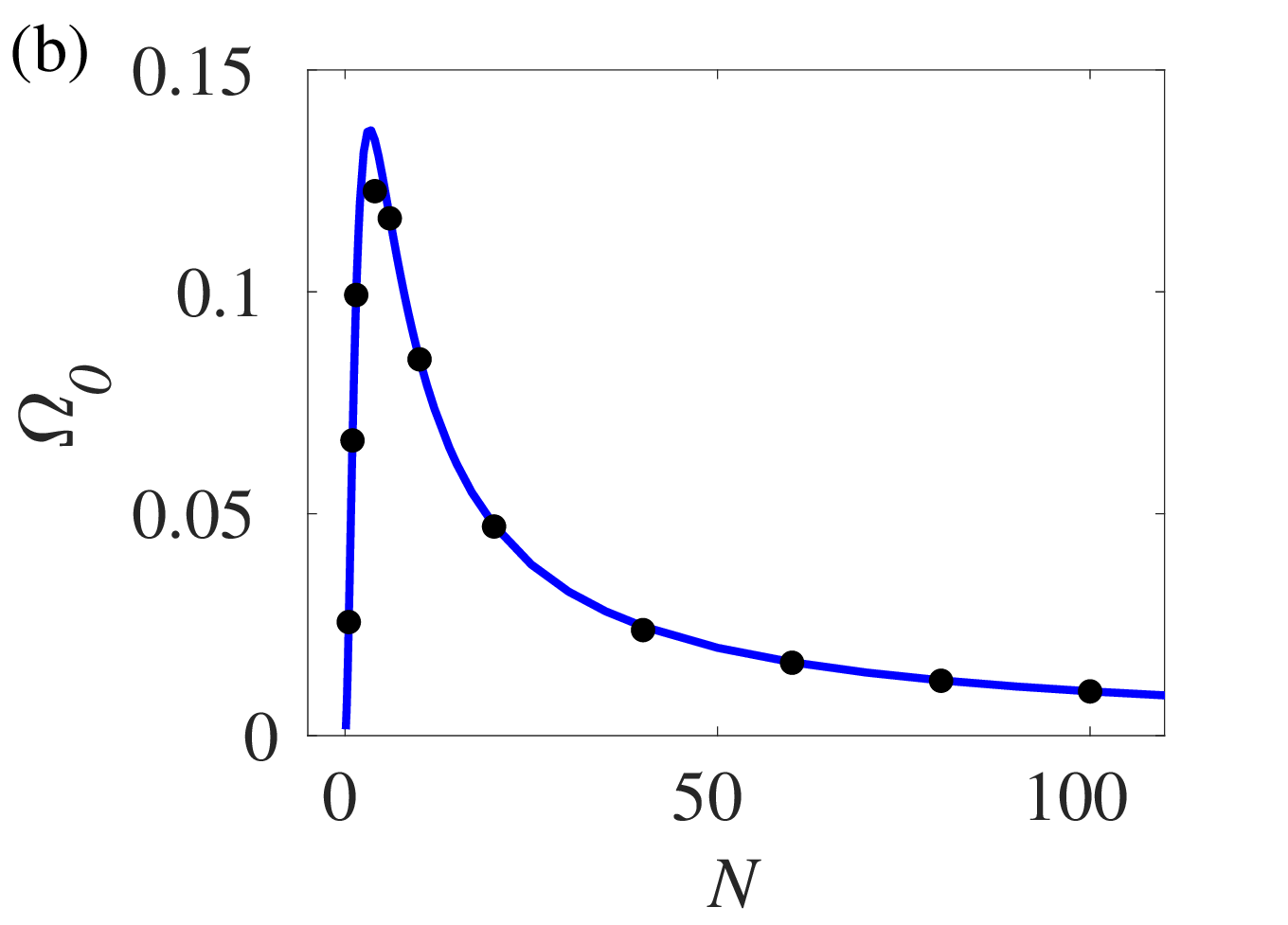}}
\caption{(a) Comparison of the QD shape oscillations specifically their amplitude and width, solid lines from numerical simulations of Eq.~(\ref{quasi_1D_gpe}) and dashed lines represent VA-predicted results for $N=10$. (b) The frequency of small oscillations, in the width of the QDs is plotted as a function of the norm $N$. This graph includes results from the VA, shown as a line, and from numerical simulations, indicated by points.  Other parameters ($\alpha, \gamma, \delta) =(0,1,1)$. }
\label{fig-3ab}
\end{figure}

Usage of the effective potential allows us to determine the frequency $\Omega_0$ associated with small oscillations of QD parameters around the equilibrium state. 
\begin{eqnarray}
& \Omega_0^2 =\left. \cfrac{\partial^2 U(w)}{\partial w^2} \right|_{w= w_s}= 2 \alpha + \cfrac{3 \Gamma(2-M)}{4 M^2 \Gamma(3M) w_s^4} 
\nonumber \\
& -\cfrac{\gamma N}{2^{M+1} M \Gamma(3M) w_s^3} +
\nonumber \\
 & \left( \cfrac{2}{5} \right)^M \cfrac{3 \delta N}{2^{5/2} M \Gamma(3M) w_s^3} \left( \cfrac{N}{\Gamma(M+1) w_s}\right)^{1/2}.
\label{omeg}
\end{eqnarray}
	In the simulations, we introduce small deviations in the VA-predicted parameters (see, Eq.~(\ref{sgauss})) and use them as initial conditions. This alteration results in the oscillation of the QD shape around the equilibrium state. Figure~\ref{fig-3ab}(a) illustrates the comparison dynamics of QD shape, specifically amplitude and width. When the amplitude reaches a maximum, the width has a minimum and vice versa, this indicates the conservation of the initial norm.  
%
We compute the average of the maximum and minimum of this small oscillation over several periods and corresponding values presented in Fig.~\ref{fig-2ab}(b) as points. This procedure allows us additional check the stability of the VA-predicted stationary solutions.

It is also evident from Fig.~\ref{fig-3ab}(b) that the frequencies predicted by the VA closely align with the numerical results for a wide range of $N$. Numerically, the frequency is determined as $\Omega_0=2 \pi / \tau$, with $\tau$ representing the average period of small oscillation.

\begin{figure}[htbp]
  \centerline{ \includegraphics[width=4.55cm]{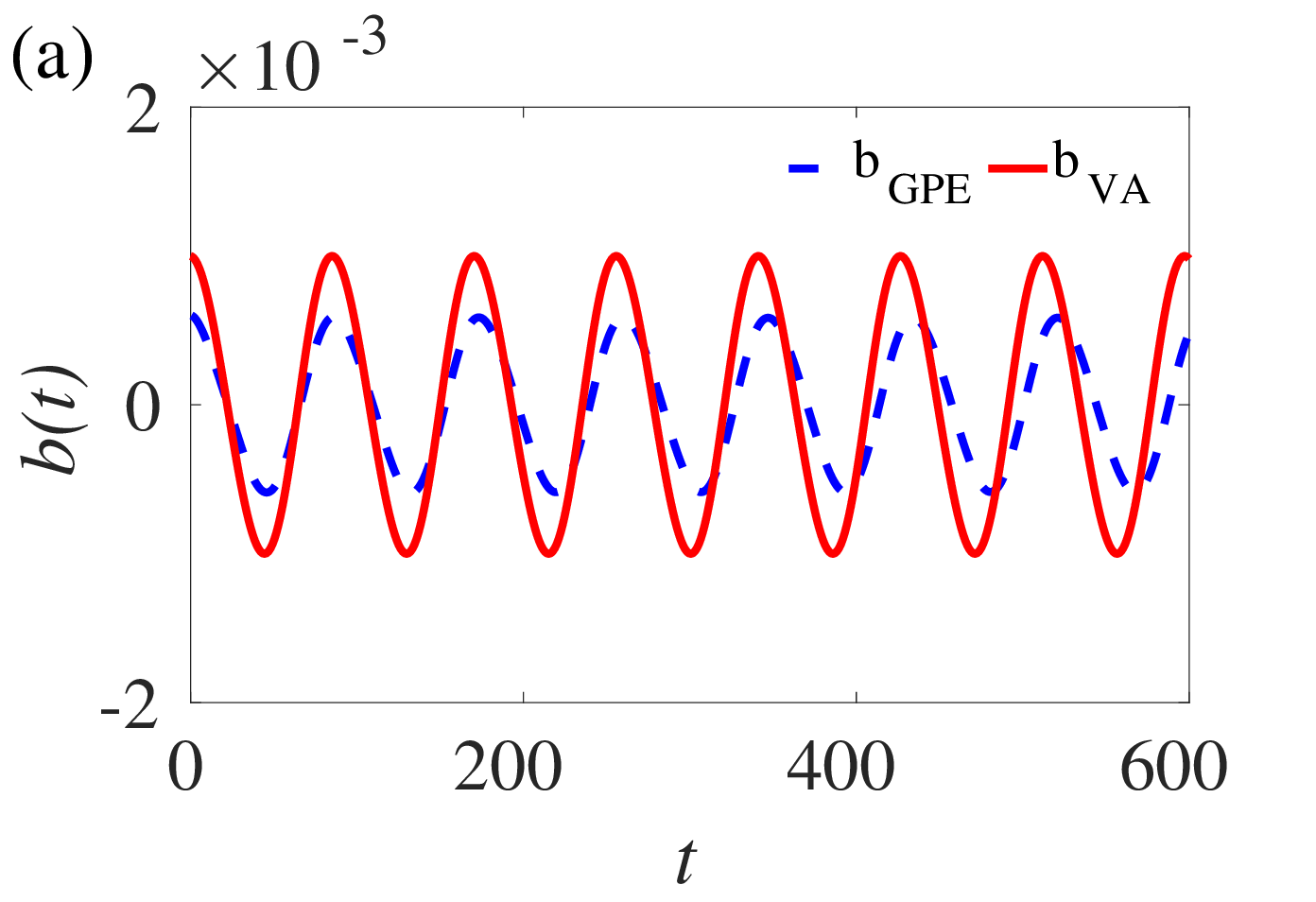} \includegraphics[width=4.55cm]{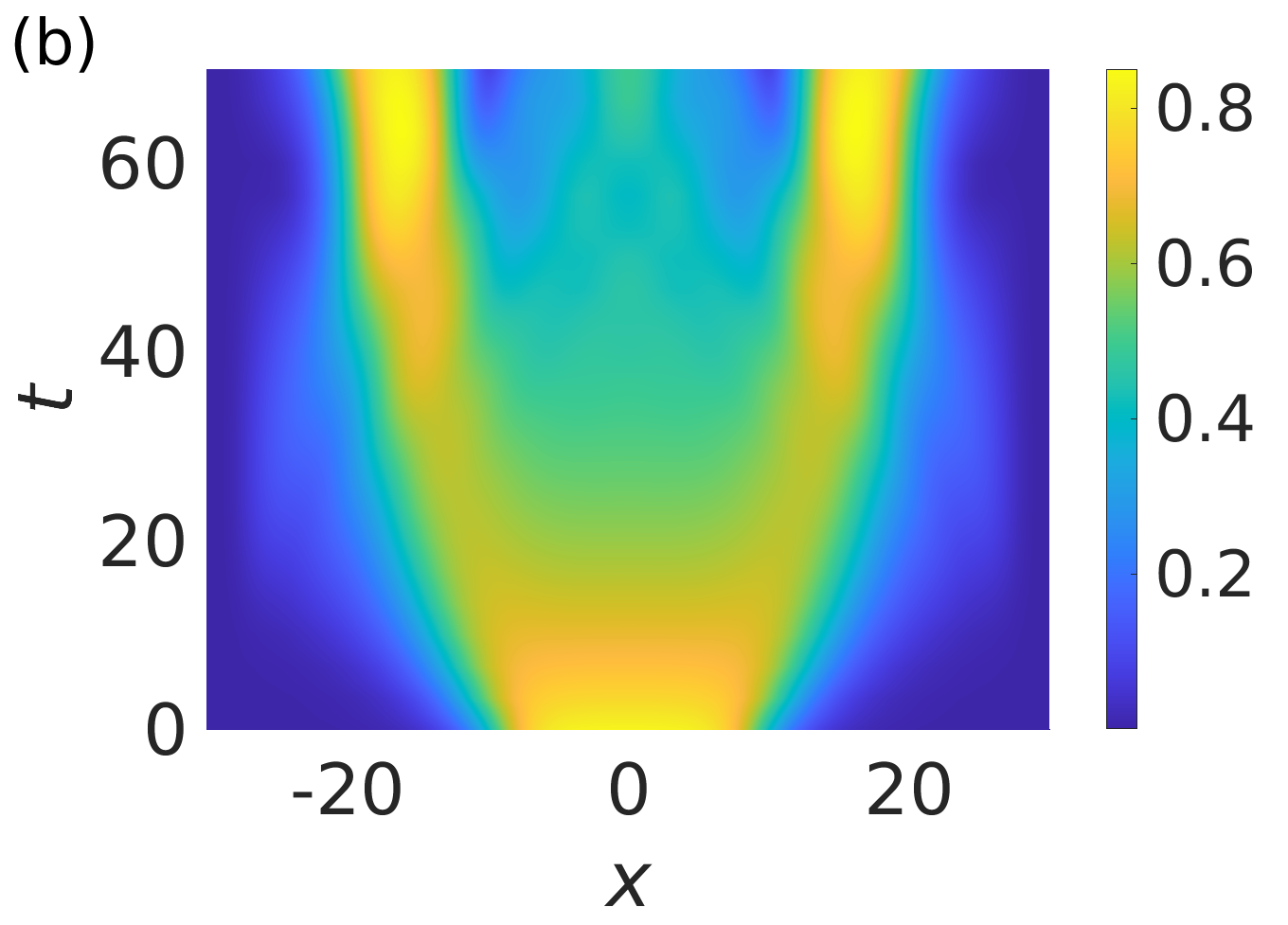}}
\caption{(a) Comparison of the chirp dynamics for the initial value $b=0.001$. The solid line represents results obtained from the VA while the dashed line corresponds to values derived from numerical simulations. (b) Dynamics of the density distributions with an initial chirp of $b=0.025$. The colorbar indicates the values of $|\Psi|$. Additional parameters are set as $(N, \gamma, \delta) = (12, 1, 1)$. } 
\label{fig-4ab}
\end{figure}

The chirp parameter allows us to describe the oscillations of the amplitude and width of the soliton in time. Also, it is the simplest extension describing the space dependence of the phase 
(see ~\cite{Michinel1997, Malomed2002, Kengne2021}).
Nonzero initial chirp in variational approximation $b_{\mathrm{VA}}$ corresponds to nonzero initial velocity $w_t $ in the equation (\ref{dyneq}) for width.  As a result, the width and chirp of the droplet start to oscillate, see Fig.~\ref{fig-4ab}(a). 
%
We compared the results for the chirp from variational equations and direct simulations. To monitor chirp in direct numerical simulations we applied the following expression~\cite{nijhof2000}
\begin{equation}
   b_{\mathrm{GPE}} =\frac{Im \int_{-\infty}^{\infty} \Psi^{* 2}_x \Psi^2 dx}{\int_{-\infty}^{\infty} |\Psi|^4dx}.
\label{Nchirp}
\end{equation}
The results of the simulations presented in Fig.~\ref{fig-4ab}(a) show that for relatively small initial chirp values, the variational equations accurately describe the evolution of droplets. However, increasing the initial chirp value leads to droplet instability, causing the initial droplet to eventually split into several fragments with some low-amplitude radiation, see Fig.~\ref{fig-4ab}(b). The variational approach cannot explain the process of splitting.

\section{Dynamics of quantum droplets under periodic management of the atomic scattering length}
\label{sec:modulation}
To illustrate the significance of the VA, we examine the periodic modulation of atomic scattering length. In experimental settings, this kind of modulation can be generated by periodic variation of the atomic scattering length using the Feshbach resonance technique~\cite{Abdullaev2023_1, Moerdijk1995, Chin2010}. Near the resonance, the scattering length as a function of the magnetic field is described by
\begin{equation}
a_s(t)=a_{bg} \left(1-\cfrac{\Delta B}{B(t) - B_r} \right), 
\label{as}
\end{equation}
where $a_{bg}$ denotes the background scattering length, characterizing the scattering length far from the resonance; $\Delta B$ represents the resonance width, showing the necessary change in the magnetic field around the resonance magnetic field $B_r$ to significantly modify the scattering length; $B_r$ is the resonance magnetic field, defined as the specific magnetic field value where the energy of the colliding atoms equals the energy of a bound state; $B(t)$ refers to the periodically varying magnetic field, which is experimentally adjusted to control the scattering length.

Assuming the magnetic field varies as $B(t)=B_0 +B_1 \sin ( \omega_{m} t)$, where $B_0$ is a constant field component, and $B_1$ is the amplitude of oscillation. Near a Feshbach resonance, this results in a corresponding variation in the scattering length $a_s(t)$, and thus in the interaction strength,~\cite{Abdullaev2003} 

\begin{equation}
\gamma(t)=\gamma_0[1+\epsilon_1 \sin(\omega_m t)], \quad \delta(t)=\delta_0[1+\epsilon_2 \sin(\omega_{m} t+\theta )],
\end{equation}
where $\epsilon_1$, $\epsilon_2 \ll 1$, $\omega_m$ and $\theta$ denote the amplitude, frequency, and initial phase of the modulations, respectively. The inclusion of $\theta$ is intended to show the impact of modulations on the initial phase difference. Initially, we set $\theta=0$. The periodic modulations in $\gamma(t)$ and $\delta(t)$ result in oscillations of the QD shape. A similar study is also reported in Ref.~\cite{Otajonov2022_1} for 3D case. 
It is also possible to study the properties of Faraday waves in the condensate using Floquet theory by periodically varying the coupling constants. However, this topic requires separate research and is beyond this paper's scope.

\begin{figure}[htbp]
  \centerline{ \includegraphics[width=4.55cm]{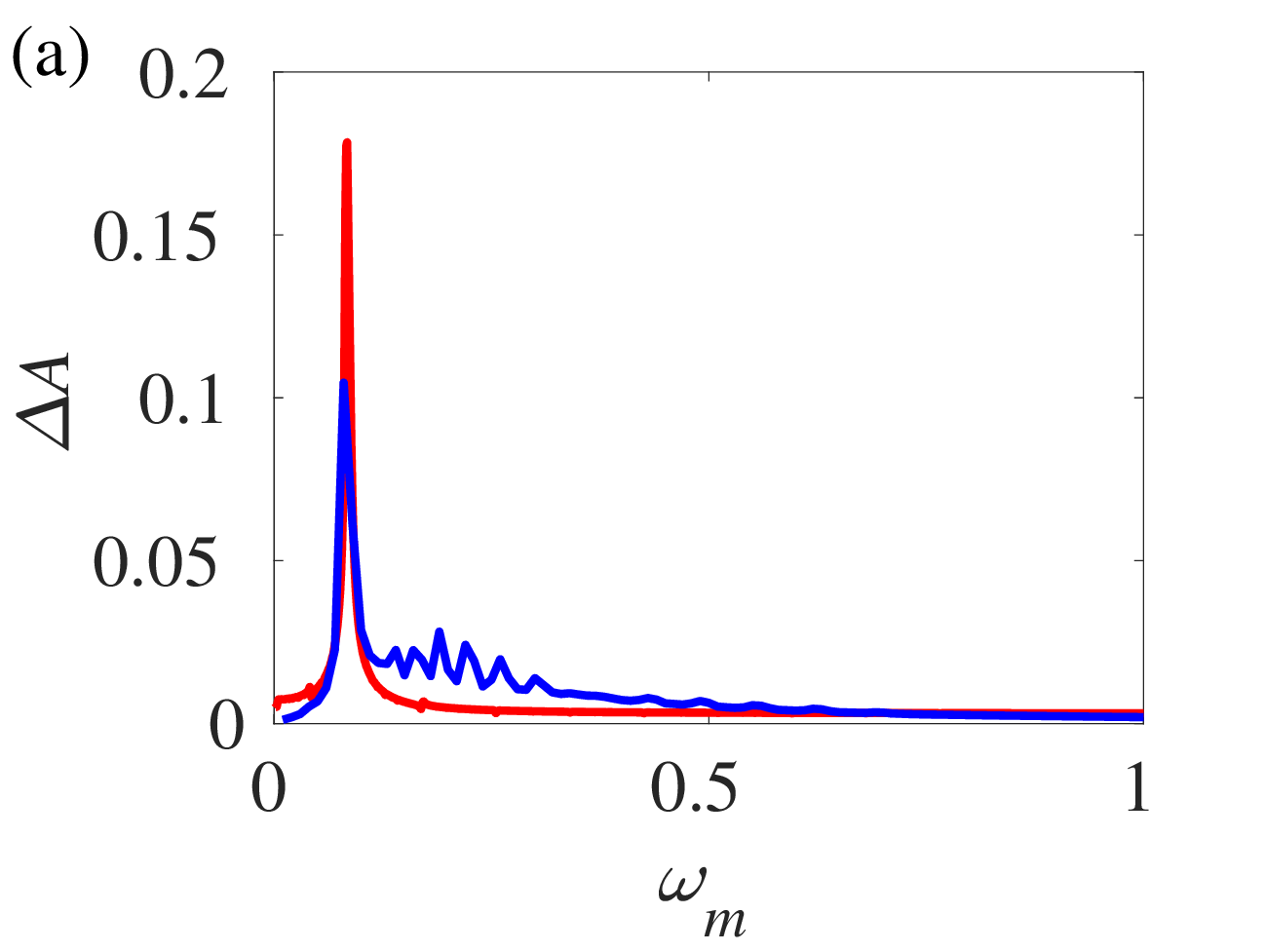} \includegraphics[width=4.55cm]{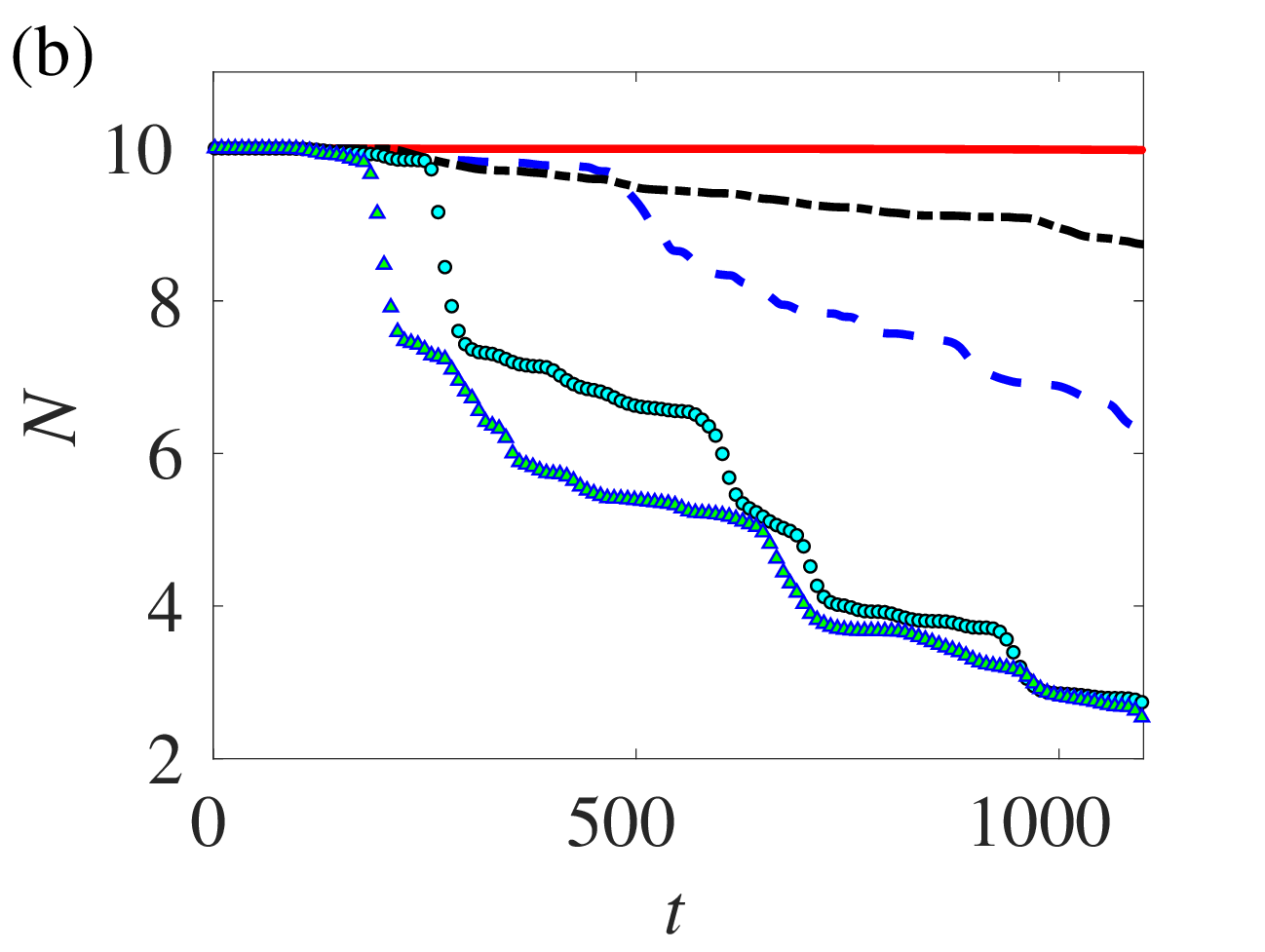}}
\caption{(a) The amplitude difference plotted against modulation frequency for $\epsilon_1 = \epsilon_2=0.05$. The line represents results obtained from numerical simulations, while points correspond to values derived from the VA. (b) The dynamics of the norm $N$ across various modulation amplitude for $\omega_m \approx 0.085$. The representations include solid lines for $\epsilon_1 = \epsilon_2 = 0.1$, dashed lines for $\epsilon_1 = 0.1, \epsilon_2 = 0.05$, and dash-dotted lines for $\epsilon_1 = 0.05, \epsilon_2 = 0.1$. Circular markers indicate the case where $\epsilon_1 = \epsilon_2 = 0.1$ and $\theta = \pi/2$, while triangular markers denote $\epsilon_1 = \epsilon_2 = 0.1$ and $\theta = \pi$. Additional parameters are set as $(N, \gamma_0, \delta_0) = (10, 1, 1)$.}
\label{fig-5ab}
\end{figure}

In simulations using Eq.~(\ref{quasi_1D_gpe}), we observe distinct responses in QD dynamics, influenced by the modulation amplitudes. The QD exhibits adiabatic oscillations for lower values of $\epsilon_1$ and $\epsilon_2$. A significant increase in oscillation amplitude occurs when the modulation frequency $\omega_m$ approaches the eigenfrequency $\Omega_0$, as shown in Fig.~\ref{fig-5ab}(a). With parameters $(\gamma_0, \delta_0)=(1,1)$, the calculated eigenfrequency $\Omega_0$ is 0.0848, while the resonance frequency $\omega_r$ identified through simulations is approximately 0.0836. This figure clearly shows that the variational method we have developed effectively determines the resonance frequencies of a quantum droplet when exposed to external periodic modulation. This resonance frequency is determined based on how the difference in amplitude $\Delta A=A_{max}-A_{min}$ varies with the modulation frequency $\omega_m$, where $A_{max}$ and $A_{min}$ represent the highest and lowest values of the amplitude oscillation over time, respectively.
Within the scenario of small periodic modulations in QD dynamics, a phenomenon akin to beating in the amplitude and width of the QD is noted, characteristic of forced oscillations. During these adiabatic oscillations, the particle count within the QD remains largely unchanged, as depicted by the solid line in Fig.~\ref{fig-5ab}(b).
%

For larger values of $\epsilon_1 =\epsilon_2$, the next type of dynamics observed involves a QD gradually diminishing as it releases particles. This emission manifests as low-density waves spreading outward from the QD. In our numerical simulations, we implemented absorbing boundary conditions to ensure that atoms exiting the quantum droplet do not reflect back from the numerical domain's edge, thereby preventing any interference with the droplet's dynamics. Furthermore, we explore scenarios where the modulation amplitudes differ, $\epsilon_1 \neq \epsilon_2$. Under these conditions, the QD decays more rapidly compared to when $\epsilon_1 = \epsilon_2$, as illustrated by the dashed and dash-dotted lines in Fig.~\ref{fig-5ab}(b).

Our findings indicate that the two forces exerted on a QD are in antiphase and show a strong dependence on $\theta$. At $\theta=0$, these forces are nearly in equilibrium, as illustrated by the solid line in Fig.~\ref{fig-5ab}(b). The most rapid decay of the QD is observed at $\theta=\pi$, as highlighted by the points in Fig.~\ref{fig-5ab}(b). This suggests that periodic modulations can serve as an efficient mechanism for altering the thermodynamic states (liquid and vapour) of BECs and for managing potential breathing mode oscillations within the QD. 

Let us expand the equation for $w_{tt}$ by a series of small deviation $\xi$ from the equilibrium width $w_s$, $w=w_s+\xi$, where $\xi \ll w_s$. The following linearized equation describes the dynamics of the width displacement $\xi$:
\begin{equation}
\ddot{\xi}+\left. {\Omega_0^2} \right|_{w=w_{s},\alpha=\alpha_0,\beta=\beta_0}\, \xi =B \sin(\omega_m t) \, ,
\end{equation}
where $\ddot{\xi}$ denotes the second order derivative of width displacement $\xi$ in terms of time $t$, and
$$B = -\cfrac{2^{-M-2} N \gamma_0 \epsilon_1}{M \Gamma(3M) w_s^2} + \cfrac{3 \cdot 2^{M-3/2} N \delta_0 \epsilon_2}{5^{M+1} M \Gamma(3 M) w_s^{5/2}} \sqrt{\cfrac{N}{\Gamma(M+1)}}\, \, . $$

\begin{figure}[htbp]
  \centerline{\includegraphics[width=6.5cm]{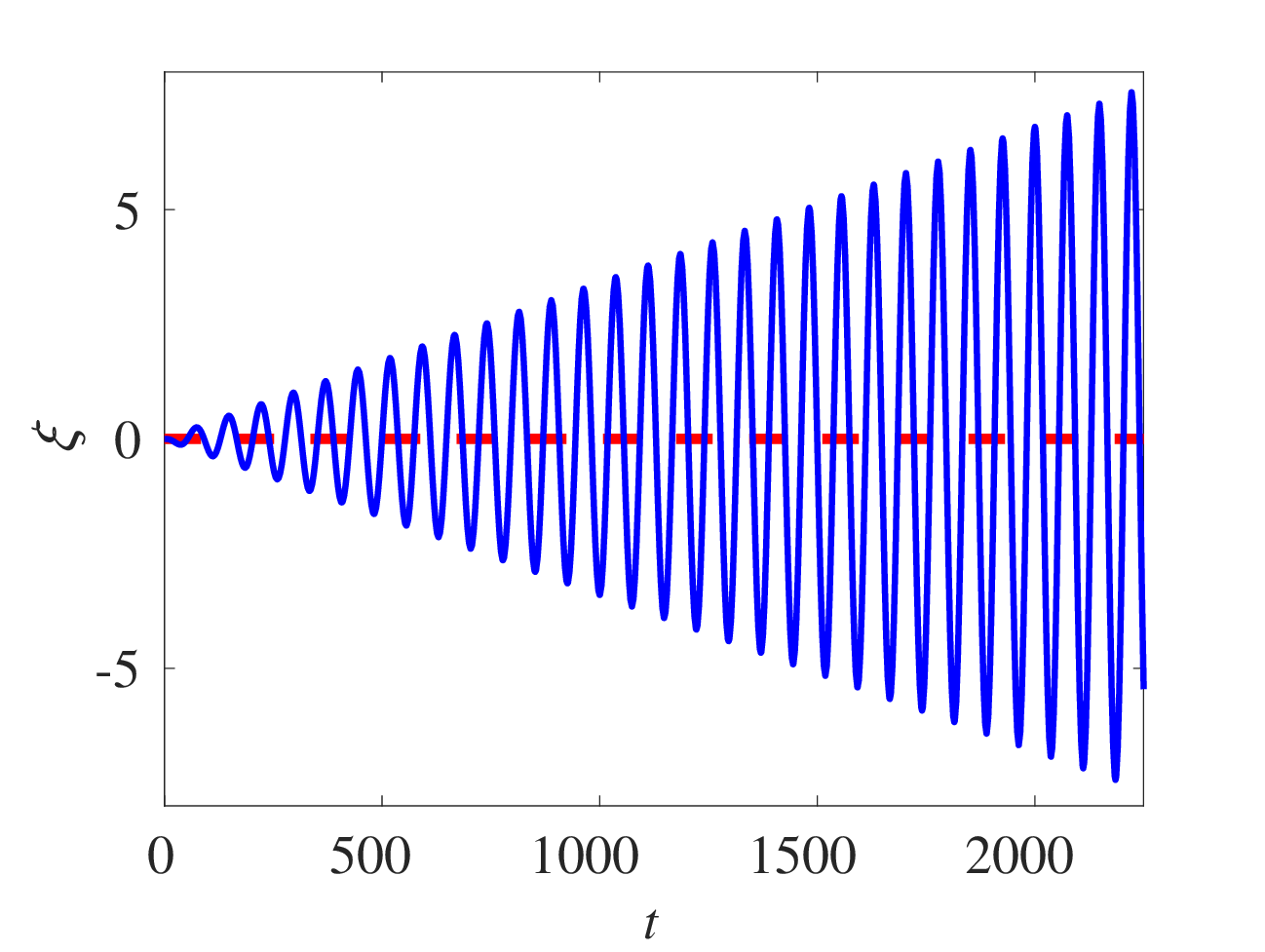} }
\caption{The behaviour of width displacement $\xi$} across various modulation amplitudes is depicted as follows: The (red) dashed line represents the case when modulation amplitude fulfils specific relation $\epsilon_1=0.1$ and $\epsilon_2=0.105641$, while the (blue) solid line indicates $\epsilon_1=\epsilon_2=0.1$. Additional parameters are set as $(\gamma_0,\delta_0,N)=(1,1,10)$.
\label{fig-6}
\end{figure}

Under the condition $B=0$, we identified the relationship between the modulation amplitudes as $\epsilon_2=5^{M+1} \gamma_0 \epsilon_1 (w_s \Gamma(M+1)) /3 \cdot 2^{2M+1/2} \delta_0 N^{1/2}$. Should the amplitude of the modulations $\epsilon_1$ and $\epsilon_2$ satisfy the given condition, their modulations cancel one another, as illustrated by the dashed line in Fig.~\ref{fig-6}. 
%
Under these circumstances, the parameters of the quantum droplets remain constant over time. The validity of this relationship is verified through the dynamics of QDs in the VA and also in numerical simulations. In the simulations, we employ an imaginary time profile as the initial condition to test this relation and note only minor, inconsequential oscillations in the QD parameters around their stationary values.
These linear analyses indicate that when the modulation amplitudes $\epsilon_1$ and $\epsilon_2$ do not meet the specified condition, the dynamics of width displacement $\xi$ are influenced by the modulation frequency $\omega_m$. Specifically, when $\omega_m$ matches the eigenfrequency $\Omega_0$, the system experiences resonant oscillations, as depicted by the solid line in Fig.~\ref{fig-6}. In proximity to the resonant frequency, a phenomenon of beating is observed. As demonstrated in Fig.~\ref{fig-6}, at the resonant frequency, the amplitude of oscillation appears to approach infinity. However, in the nonlinear dynamics at the resonant frequency, there is a noticeable beating effect in the QD parameters, attributable to the interdependence of oscillation amplitude and frequency in a nonlinear system.

\section{Lee-Huang-Yang fluid}
\label{sec:LHYfluid}

	In this section, our attention is directed toward the LHY fluid~\cite{Jorgensen2018, Skov2021}. By manipulating scattering lengths through the Feshbach resonance technique and adjusting atom numbers within a binary BEC, it becomes feasible to create a scenario where attractive and repulsive two-body interactions offset each other. In this case, the residual mean-field interactions vanish from the governing equation, with $\delta a = 0$, implying that $a$ should equal $a_{12}$ when $\delta a = 0$, the scale parameters must be redefined: 
$$t_s=\omega_{\perp}^{-1}\,, \quad
x_s=\sqrt{\cfrac{\hbar}{m_0 \omega_{\perp}}}\,, \quad \psi_s=\left(\cfrac{15 \pi m_0 l_{\perp}^3 \omega_{\perp} \delta}{512 \sqrt{2} \hbar a^{5/2}} \right)^{1/3} \, .$$
	In this system, repulsive quantum fluctuations take precedence as the dominant interaction term. When we balance these repulsive quantum fluctuations with an external potential, it becomes possible to achieve a state known as the LHY fluid.
	
\begin{figure}[htbp]
  \centerline{ \includegraphics[width=4.55cm]{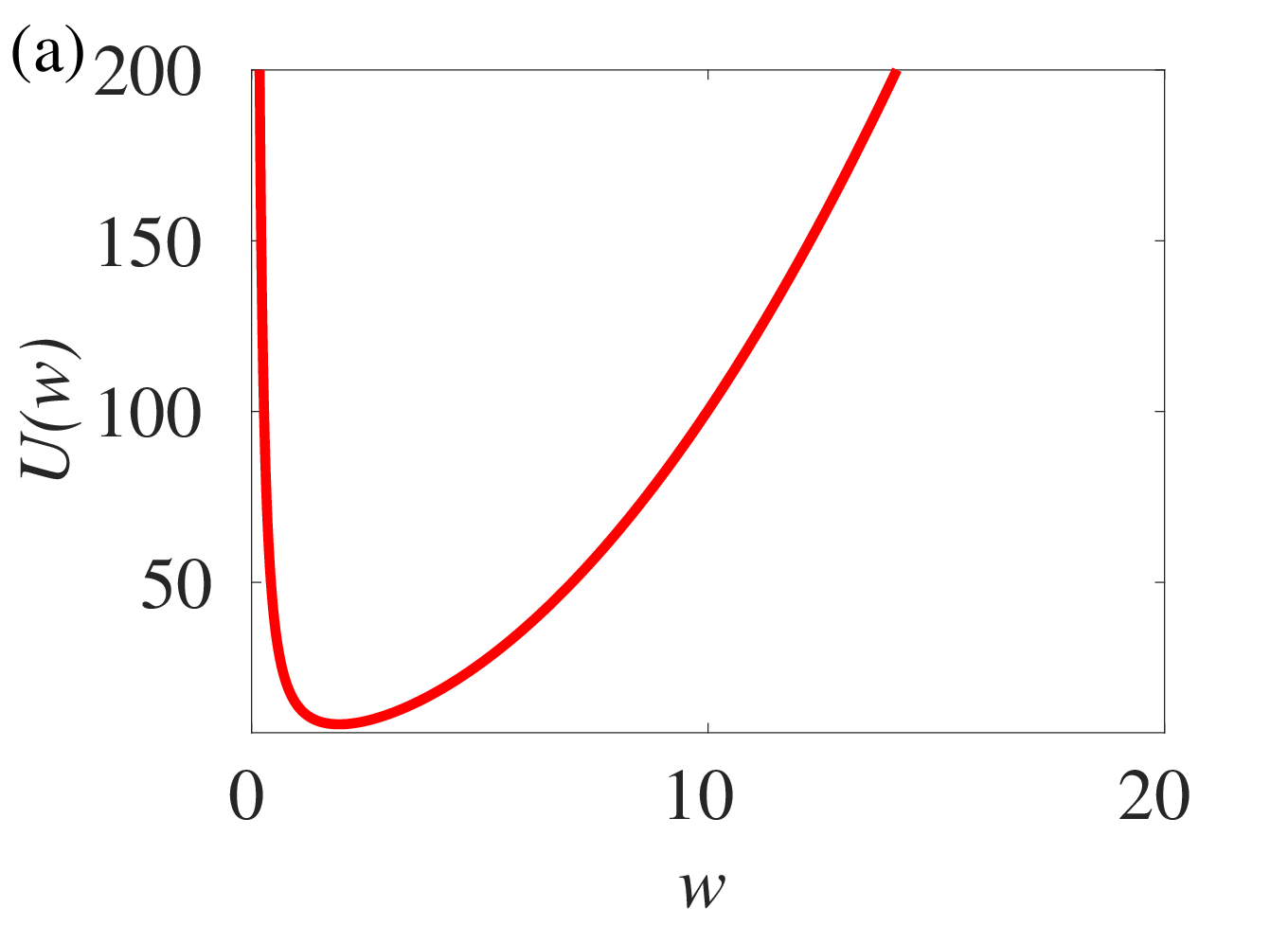} \includegraphics[width=4.55cm]{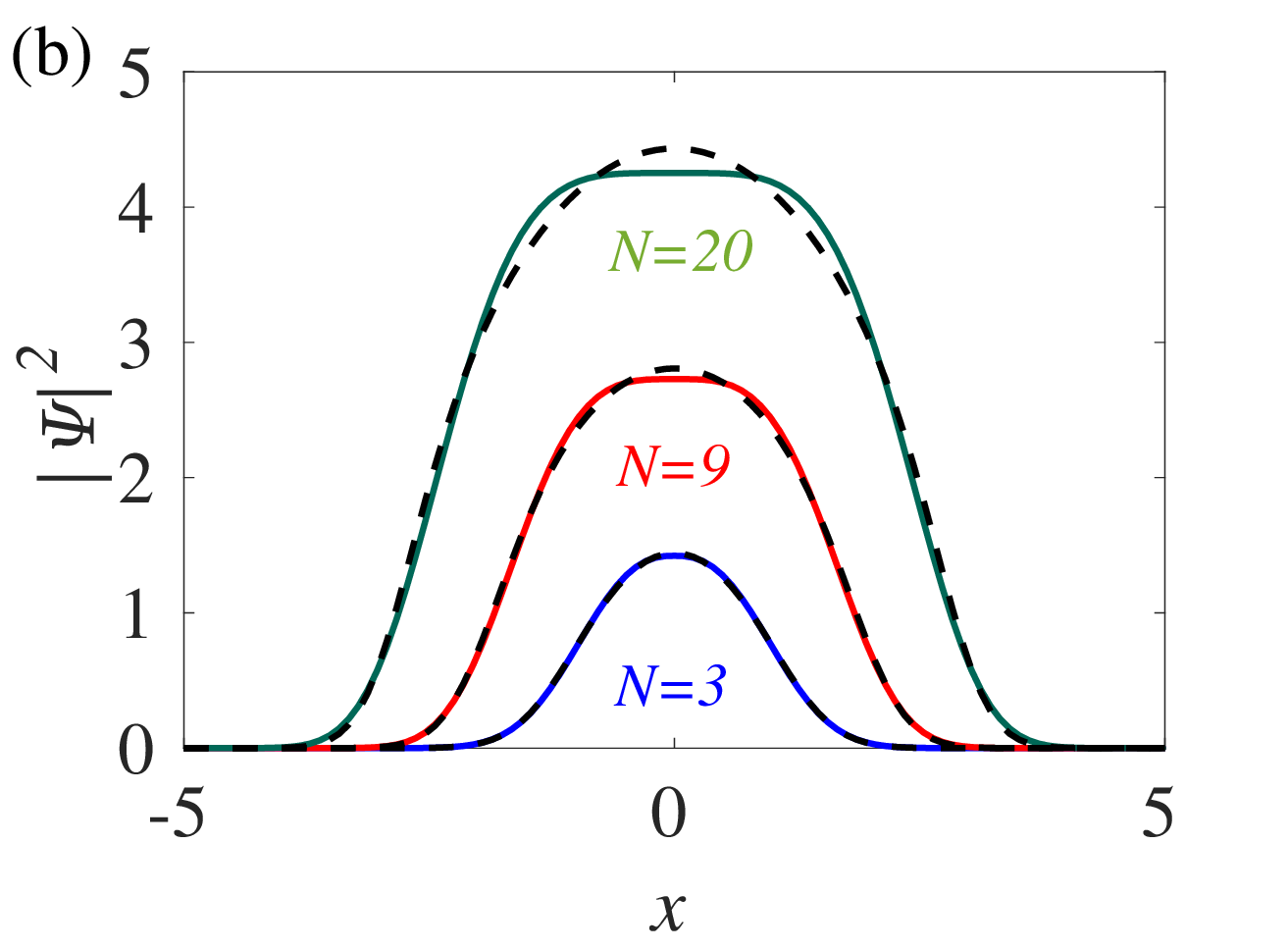}}
\caption{(a) Typical shape of the potential curves $U(w)$ for $N=10$. (b) The stationary QDs profiles correspond to different particle numbers, where solid lines represent results derived using the VA, and dashed lines are outcomes from imaginary time simulations. Other parameters are  ($\alpha, \gamma, \delta) =(1,0,1)$.}
\label{fig-7ab}
\end{figure}

In our model, $\alpha \ne 0$ and $\gamma=0$ correspond to the LHY fluid case. In this case, the characteristic form of the effective potential is illustrated in Fig.~\ref{fig-7ab}(a). In the scenario where $\alpha=0$ and $\gamma \neq 0$, the potential curves tend to approach zero as $w$ becomes large, as depicted in Fig.~\ref{fig-1ab}(a). In contrast, in the LHY fluid case, the potential takes on a parabolic form, as shown in Fig.~\ref{fig-7ab}(a), see also Eq.~(\ref{pot}). In both cases, the minimum of the potential curves corresponds to the stationary width of the localized states. 
%

In figure~\ref{fig-7ab}(b), we compared the stationary profiles of the LHY fluid predicted by the VA and the numerical simulated solutions for different values of the number of atoms. In the case of the LHY fluid, it is evident that as the number of atoms increases, the density of the localized state does not reach saturation. Conversely, in the scenario where $\alpha=0$ and $\gamma \neq 0$, the localized state exhibits a saturation density at large values of $N$, indicative of incompressible liquid properties, as depicted in Figs.~\ref{fig-1ab}(b) and ~\ref{fig-2ab}(b). In the LHY fluid case, a thorough investigation across a wide range of $N \in (0, 2000]$ did not reveal a saturation density. 

\begin{figure}[htbp]
  \centerline{ \includegraphics[width=4.55cm]{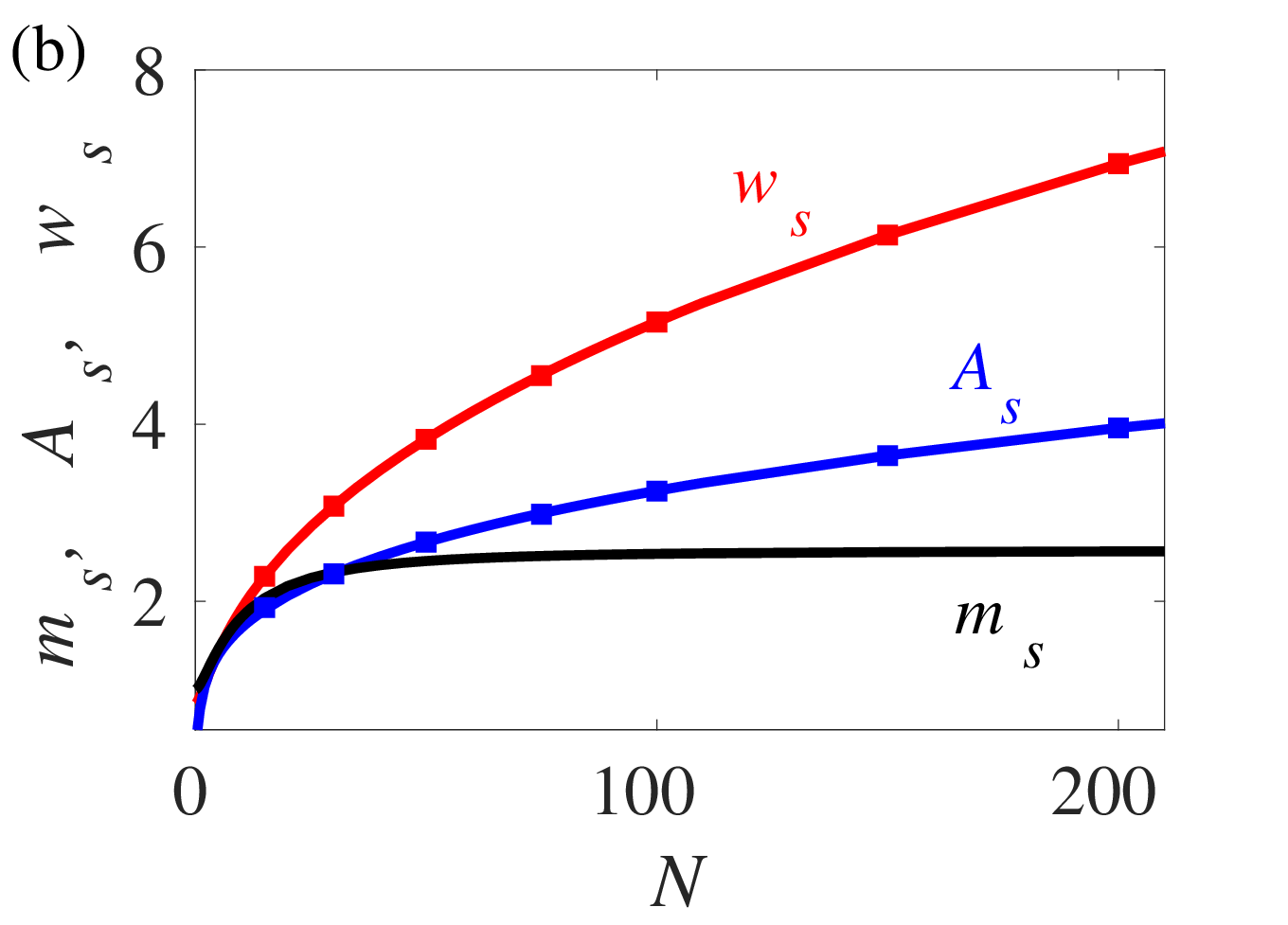} \includegraphics[width=4.55cm]{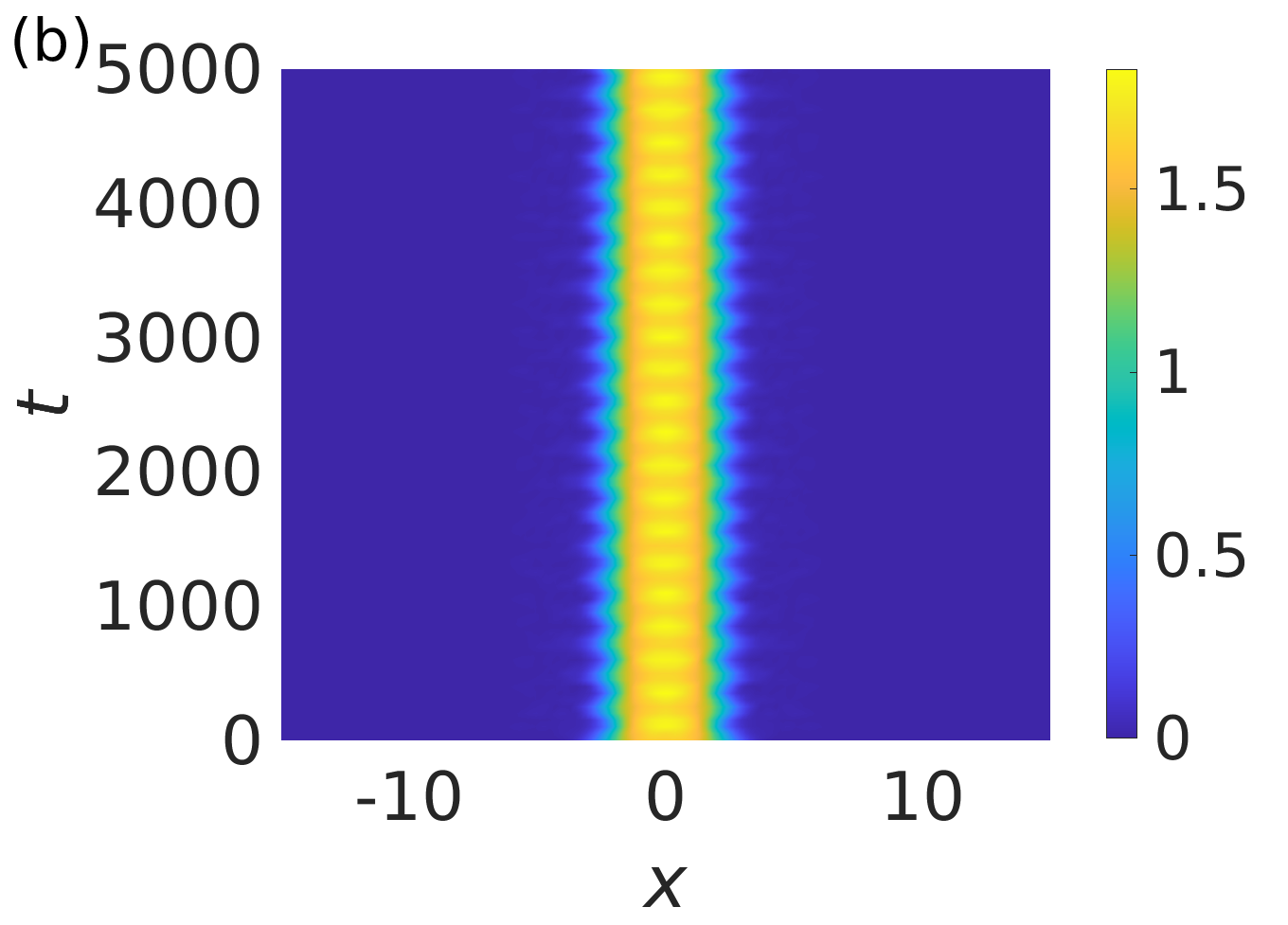}}
\caption{(a) The steady-state parameters for the LHY fluid at different values of $N$, where straight lines are determined through the VA, while points are obtained from numerical simulations. (b) The dynamics of $|\Psi|^2$ of the initially perturbed LHY fluid for $N=10$. Other parameters are  ($\alpha, \gamma, \delta) =(1,0,1)$.}
\label{fig-8ab}
\end{figure}

In figure~\ref{fig-8ab}(a), a comparison is presented regarding the stationary parameters of LHY fluid. The lines found from the VA, and points represent the numerical simulated averaged values. With an increase in the number of atoms $N$, both the amplitude and width of the LHY fluid exhibit a corresponding increase. Additionally, the super-Gaussian indices $m_s$ also experience an extremely slow increment on large $N$. The stationary profile of the LHY fluid also transforms into a more flat-top shape as $N$ increases. In the simulations, a small initial perturbation is introduced in the VA-predicted parameters, providing an additional check on the stability of the solutions, as discussed after Eq.~(\ref{omeg}). We verified the stability of localized states across a broad range of atom numbers, and in all observed cases, stable evolutions were confirmed. Figure~\ref{fig-8ab}(b) depicts the typical stable time evolution of the initially perturbed localized states.

Let us determine the parameters of our model for realistic experimental conditions. We consider $^{39}\mathrm{K}$ atoms in different spin states, for example $|F=1,m_F=1\rangle$ and $|F=1, m_F=0\rangle$, with a mass of $m_0 = 6.49 \times 10^{-26}$ kg. Both intra- and inter-species scattering lengths are set as $a_{11} = a_{22} = a = 50\,a_0$ and $a_{12} = -50 \,a_0$, where $a_0$ is the Bohr radius, ensuring that the residual scattering length satisfies $\delta a = 0$. The characteristic scales of our system, with $|\delta| = 1$, are $x_s = l_\perp \approx 0.7$ $\mu$m, $t_s \approx 0.3$ ms, and $\psi_s \approx 5.1 \times 10^4$ $m^{-1/2}$. We chose a transverse trap frequency of $\omega_\perp = 2 \pi \times 500$ Hz, providing tight confinement and allowing the condensate dynamics to be considered within a quasi-1D approximation. The integration domain length $L = 500$ corresponds to 360 $\mu$m, while dimensionless $t = 1000$ equates to $300$ ms. Real atom numbers and the size of the QD corresponding to $N = 10$ are $N_{\mathrm{real}} = 1.8 \times 10^4$ and $5.7$ $\mu$m, respectively. The critical temperature for these values is $T_{\mathrm{cr}} = 3.3125 \cfrac{\hbar^2}{m_0 k_B} \cdot \left( \cfrac{N_{\mathrm{real}}}{V} \right)^{2/3} \approx 0.88 \, \mu K$, where $N_{\mathrm{real}} = N \psi_s^2 x_s$ and $V = L x_s l_{\perp}^2$. 
The estimates provided above demonstrate that the parameter values used in numerical simulations qualitatively align with experimental conditions.

\section{Collision dynamics of quasi-1D quantum droplets}
\label{sec:collision}

In our model, the condensate wave function, as defined by Eq.~(\ref{sgauss}), is structured by the interplay of cubic and quartic nonlinear terms and dispersion within the governing GPE for the QD. This interplay generates a form reminiscent of the typical bright solitons. It is a well-established fact that the shape of the density profile remains unchanged during its time evolution, even when it is in motion at a steady velocity.
One distinctive feature of solitons in integrable systems is their ability to maintain their shape during collisions with other solitons. When $\alpha=0$ and $\delta=0$, Eq.~(\ref{quasi_1D_gpe}) transforms into a 1D nonlinear Schr\"{o}dinger equation, which is an integrable equation. In this scenario, collisions between solitons are entirely elastic. 
From this viewpoint, in our quasi-1D model, it becomes essential to investigate the persistence of the shape of droplets engaged in pairwise collisions. 
A similar study is reported in Ref.~\cite{Astrakharchik2018} for the 1D case with a quadratic-cubic nonlinear term. In the present Section, to investigate the collision dynamics we numerically solved Eq.~(\ref{quasi_1D_gpe}) by using the split-step Fourier method with periodic boundary conditions and VA-predicted initial conditions, see Eq.~(\ref{sgauss}).
The initial condition was taken as a superposition of two QDs,
\begin{equation}
\Psi(x,t=0)=\Psi_1(x+x_0) \mathrm{e}^{i k x}+\Psi_2(x-x_0) \mathrm{e}^{-i k x +i \beta}  
\label{superpozition}
\end{equation}
here, $\Psi_1(x)$ and $\Psi_2(x)$ represent the stationary shapes of QD normalized with $N_1$ and $N_2$, as derived from Eq.~(\ref{sgauss}). The parameters $\pm x_0$ denote their initial positions, $\pm k$ represent the initial momenta of the colliding droplets, and $\beta$ stands for the relative phase. 

\begin{figure}[htbp]
  \centerline{ \includegraphics[width=4.55cm]{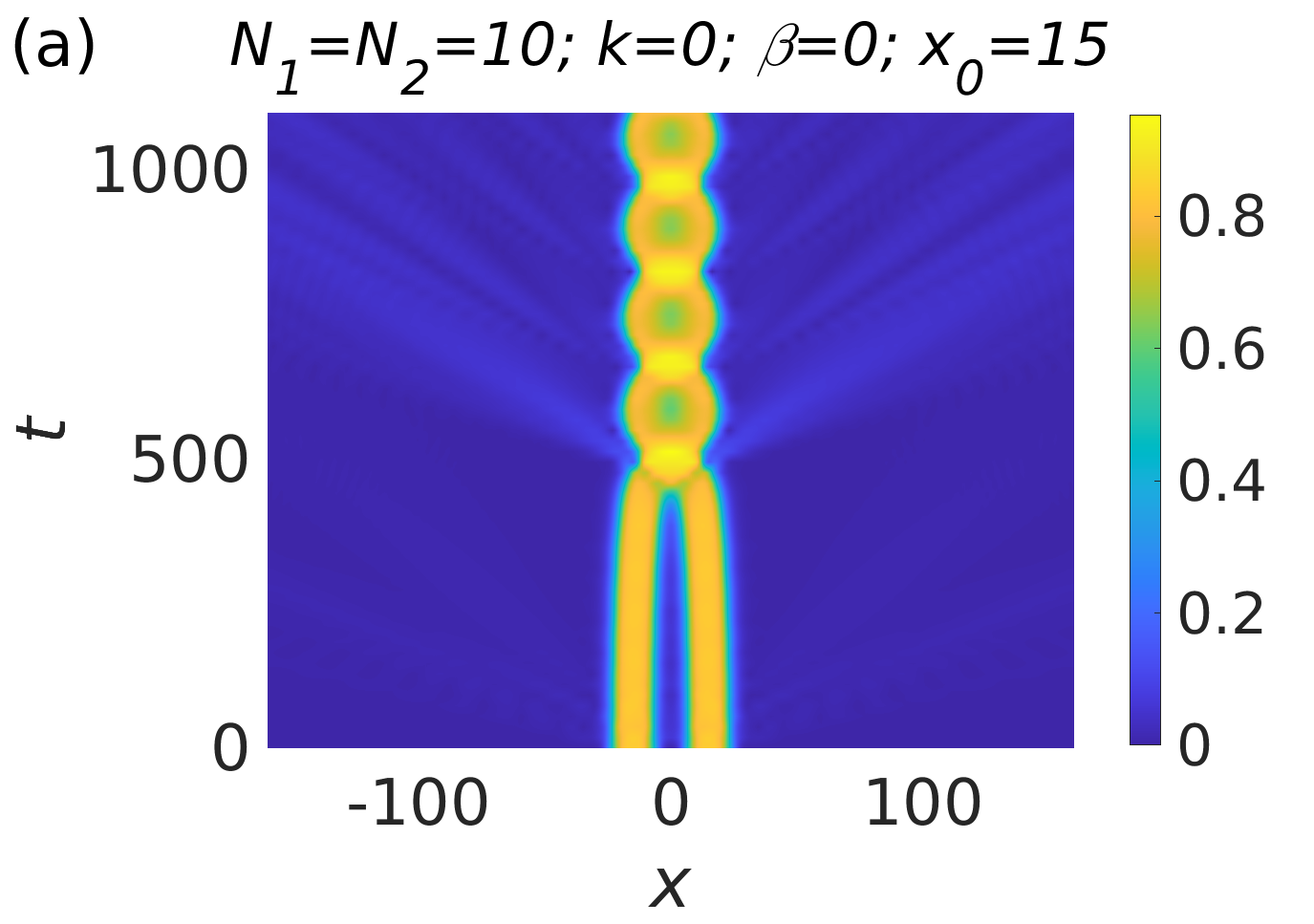} \includegraphics[width=4.55cm]{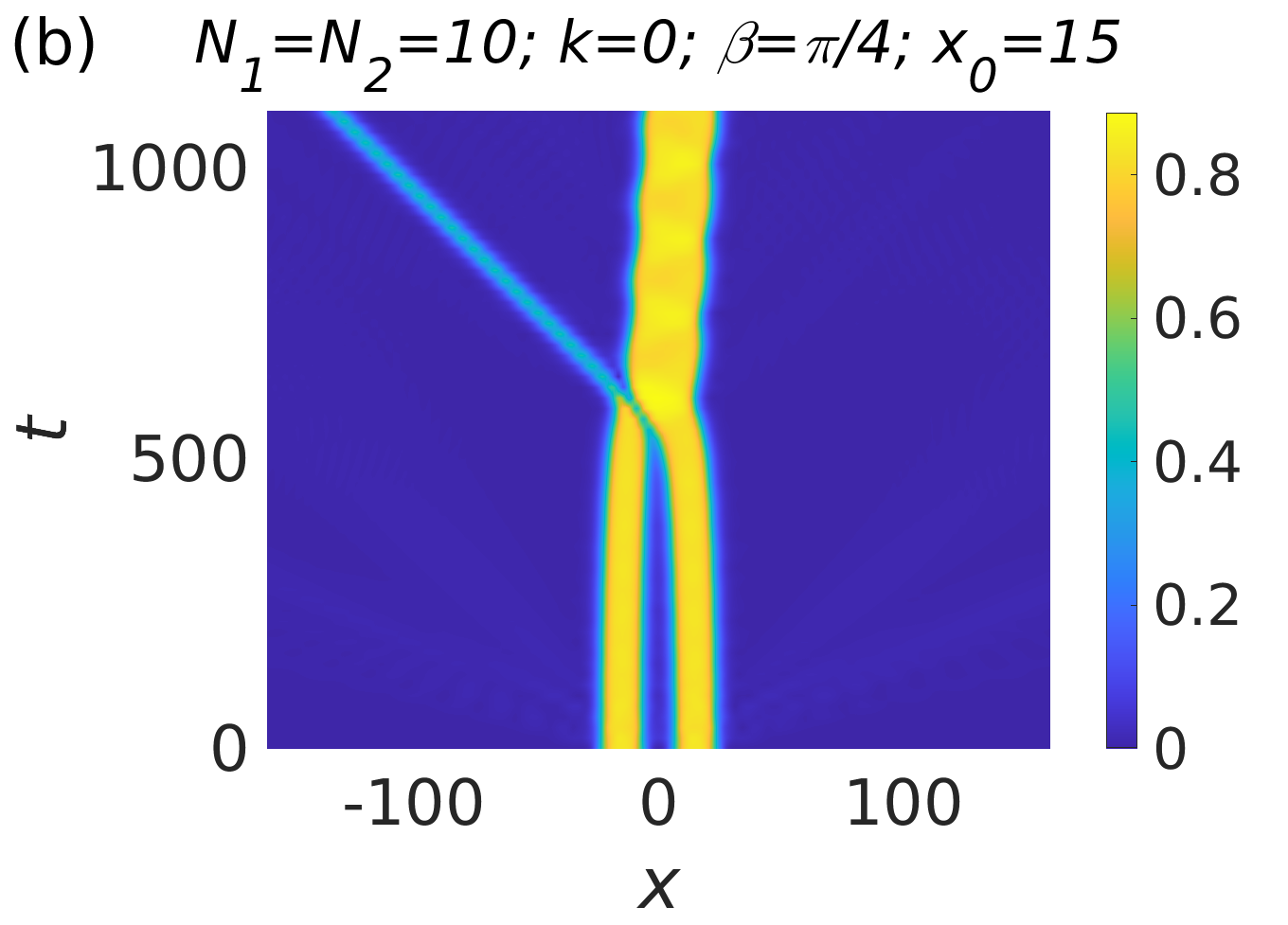}}
  \centerline{\includegraphics[width=4.55cm]{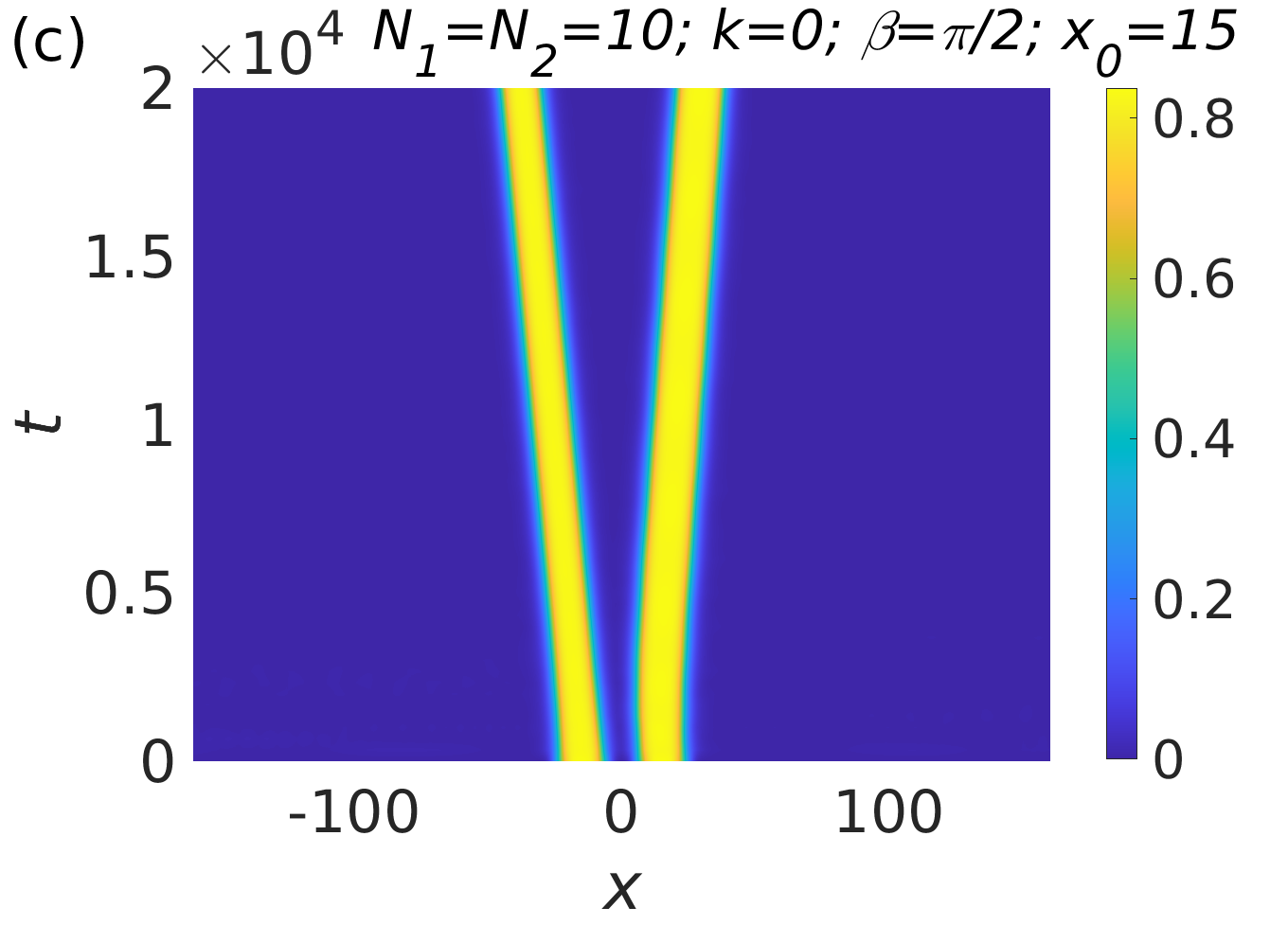} \includegraphics[width=4.55cm]{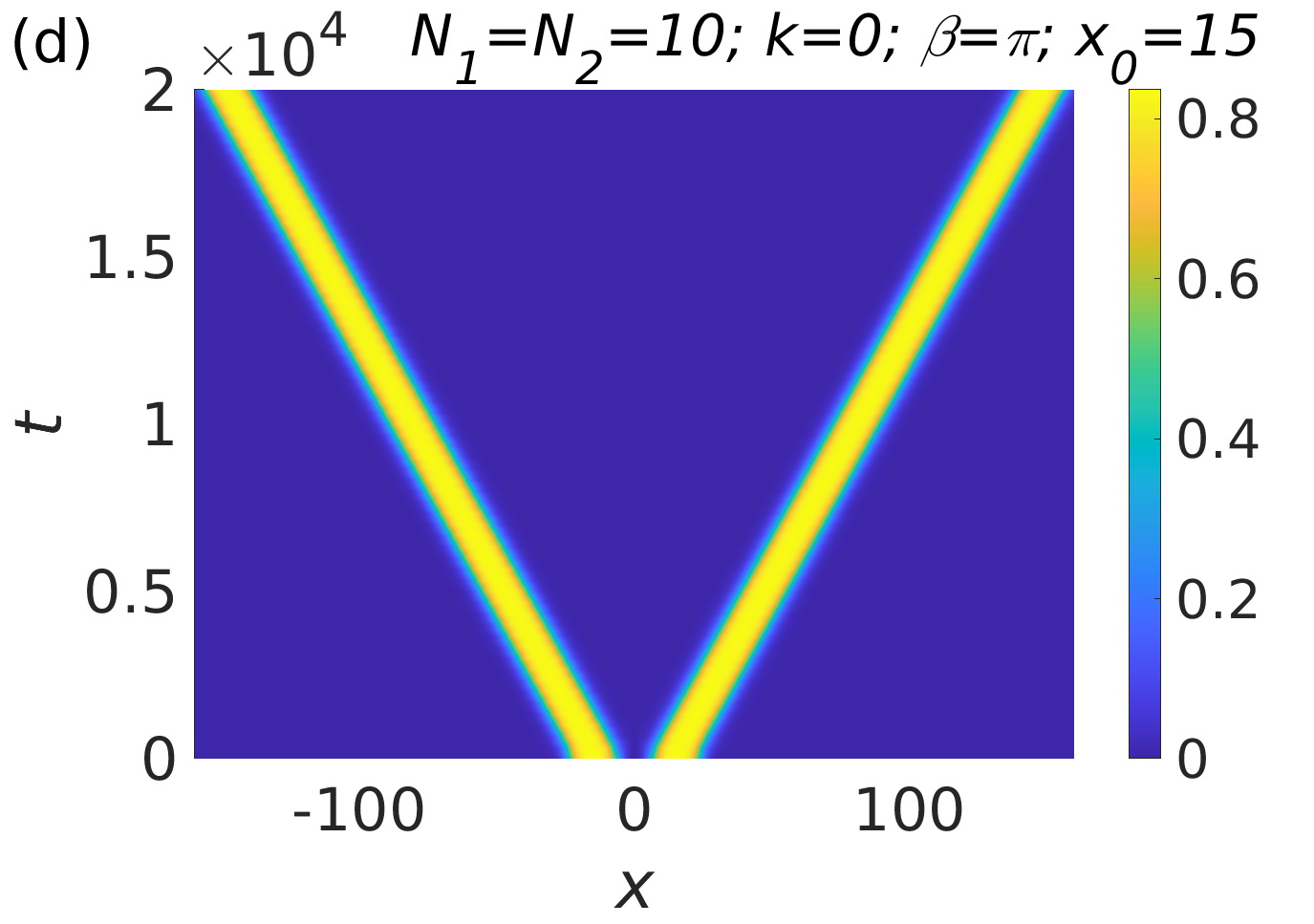}}
\caption{Density plots illustrate the evolution of two identical stationary droplets for different relative phase values $\beta$. These include (a) in-phase with $\beta=0$, (b) a phase of $\beta=\pi/4$, (c) a phase of $\beta=\pi/2$, and (d) out-of-phase with $\beta=\pi$.}
\label{fig-9abcd}
\end{figure}

Let us start with the interaction of two non-moving QDs. When two stationary QDs are positioned at a considerable distance from each other, they do not interact. It is only when they are brought close enough, such that the ``tails" of the localized waves overlap, that they begin to interact. 
In this scenario, it is observed that the interaction force between two identical QDs is contingent upon their relative phase, as depicted in Fig.~\ref{fig-9abcd}. 

When two in-phase ($\beta=0$) droplets are present, they attract each other and merge, as illustrated in Fig.~\ref{fig-9abcd}(a). At the collision point, there might be radiation of small amplitude waves. An asymmetric interaction is noted between two identical quantum droplets when the relative phase falls within the interval $0<\beta \le \pi/2$. The interaction dynamics in these regions reveal two inelastic regimes, as seen in Figs.~\ref{fig-9abcd}(b) and (c). In Fig.~\ref{fig-9abcd}(b), two QDs collide with a phase difference of $\pi/4$, initially attracting each other, colliding, and then repelling, with particle redistribution at the collision point. In Fig.~\ref{fig-9abcd}(c), two QDs interaction is presented with a phase difference of $\pi/2$, where redistribution occurs through atom exchange during the initial interval, followed by repulsion. The resulting QDs maintain their shape during motion, and their velocities differ to comply with momentum conservation. In the $\pi/2 <\beta \le \pi$ interval, QDs induce repulsion, with maximal repulsion occurring at $\beta=\pi$. The initial density distribution is also preserved, as shown in Fig.\ref{fig-9abcd}(d).

\begin{figure}[htbp]
\centerline{ \includegraphics[width=4.55cm]{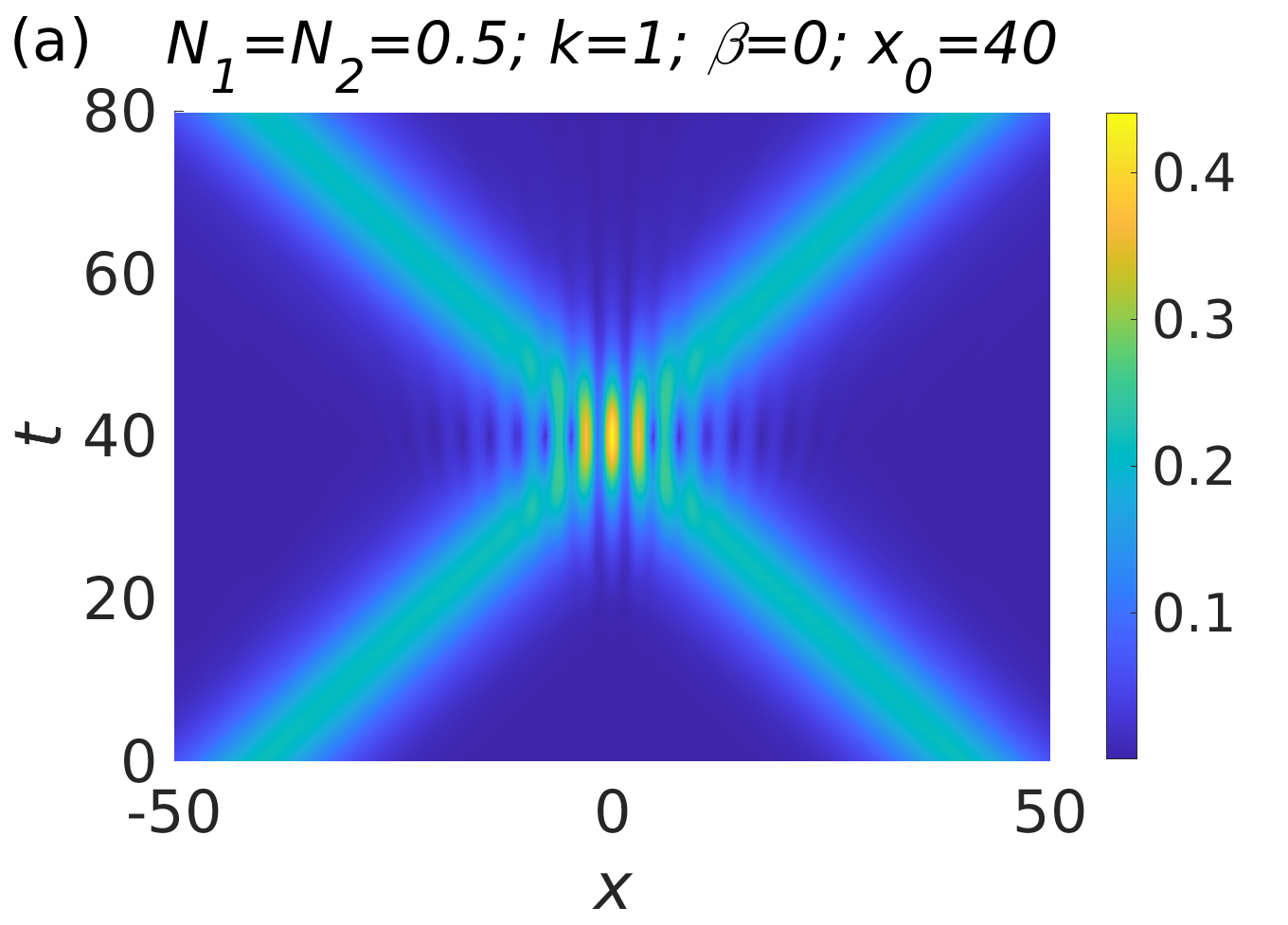} \includegraphics[width=4.55cm]{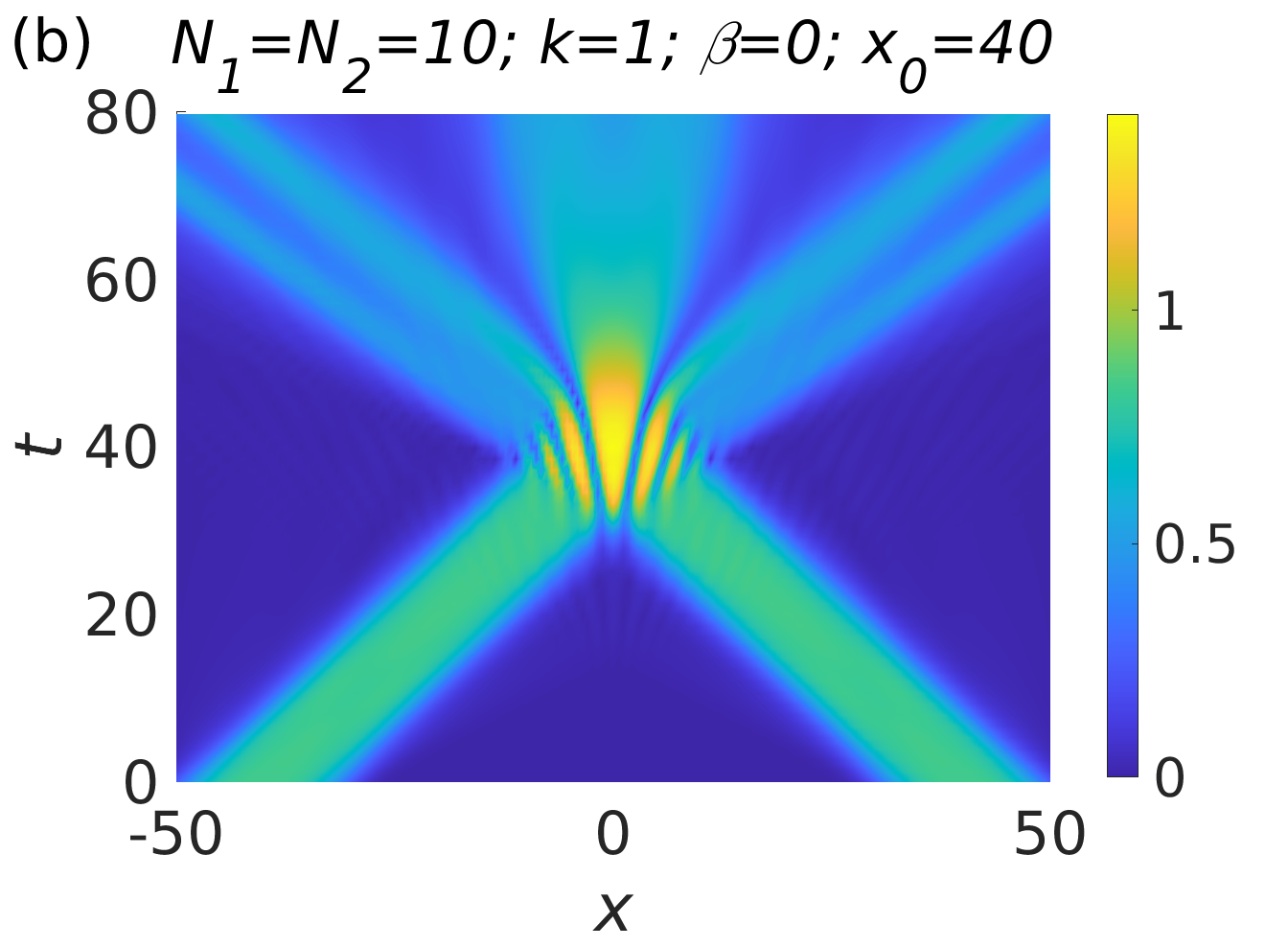}}
\caption{The time evolution density plot reveals an interference pattern resulting from the interaction of two identical, in-phase ($\beta = 0$) droplets, each having an incident momentum of $k=1$. This is observed (a) in the case of small $N_1=N_2=0.5$ bell-shaped QDs, and (b) for larger $N_1=N_2=10$ flat-top QDs. Other parameters are  ($\alpha, \gamma, \delta) =(0,1,1)$, and $x_0=40$.}
\label{fig-10ab}
\end{figure}

Now, let's explore the collision dynamics between two quantum droplets travelling in opposite directions. The interaction force between these QDs is generally influenced by factors such as the phase difference, initial momentum (velocity), and the number of particles. Figure~\ref{fig-10ab}(a) illustrates the collision dynamics of fast-moving bell-shaped droplets with a small number of particles, while Fig.~\ref{fig-10ab}(b) depicts the collision dynamics of flat-top QDs with a large number of particles. In both cases, an interference pattern may manifest at the moment of collision, serving as a distinctive feature of the interplay of coherent matter waves. 

For small bell-shaped quantum droplets with $N_1 = N_2 = 0.5$ and $k=1$, the density distribution of the droplets remains unchanged after the collision. However, in the case of large flat-top QDs, the shape of the droplet is not preserved post-collision. Instead, interactions between the incoming droplets lead to the formation of multiple outgoing small-amplitude droplets, along with additional quiescent droplets.

\begin{figure}[htbp]
  \centerline{ \includegraphics[width=4.55cm]{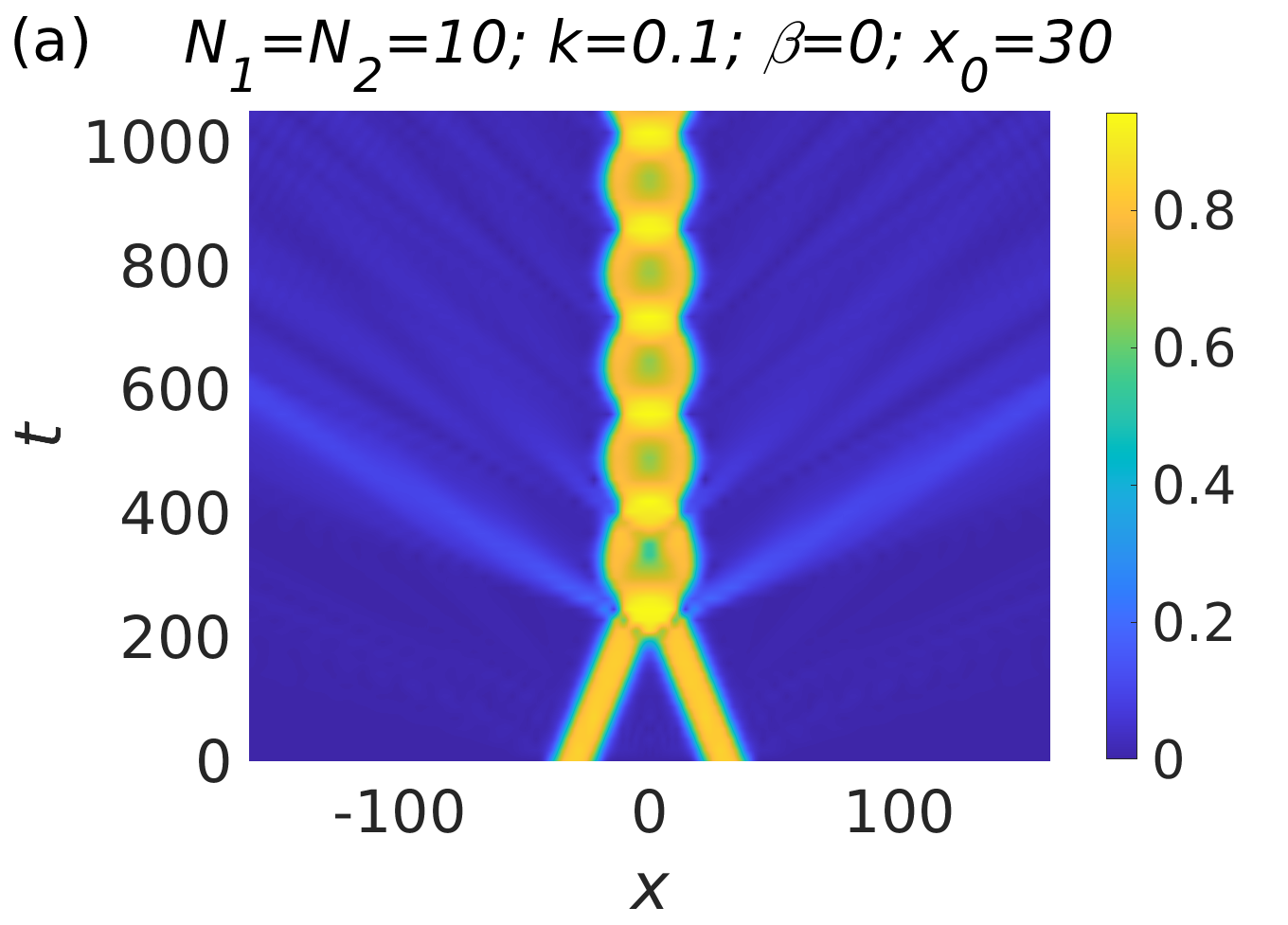} \includegraphics[width=4.55cm]{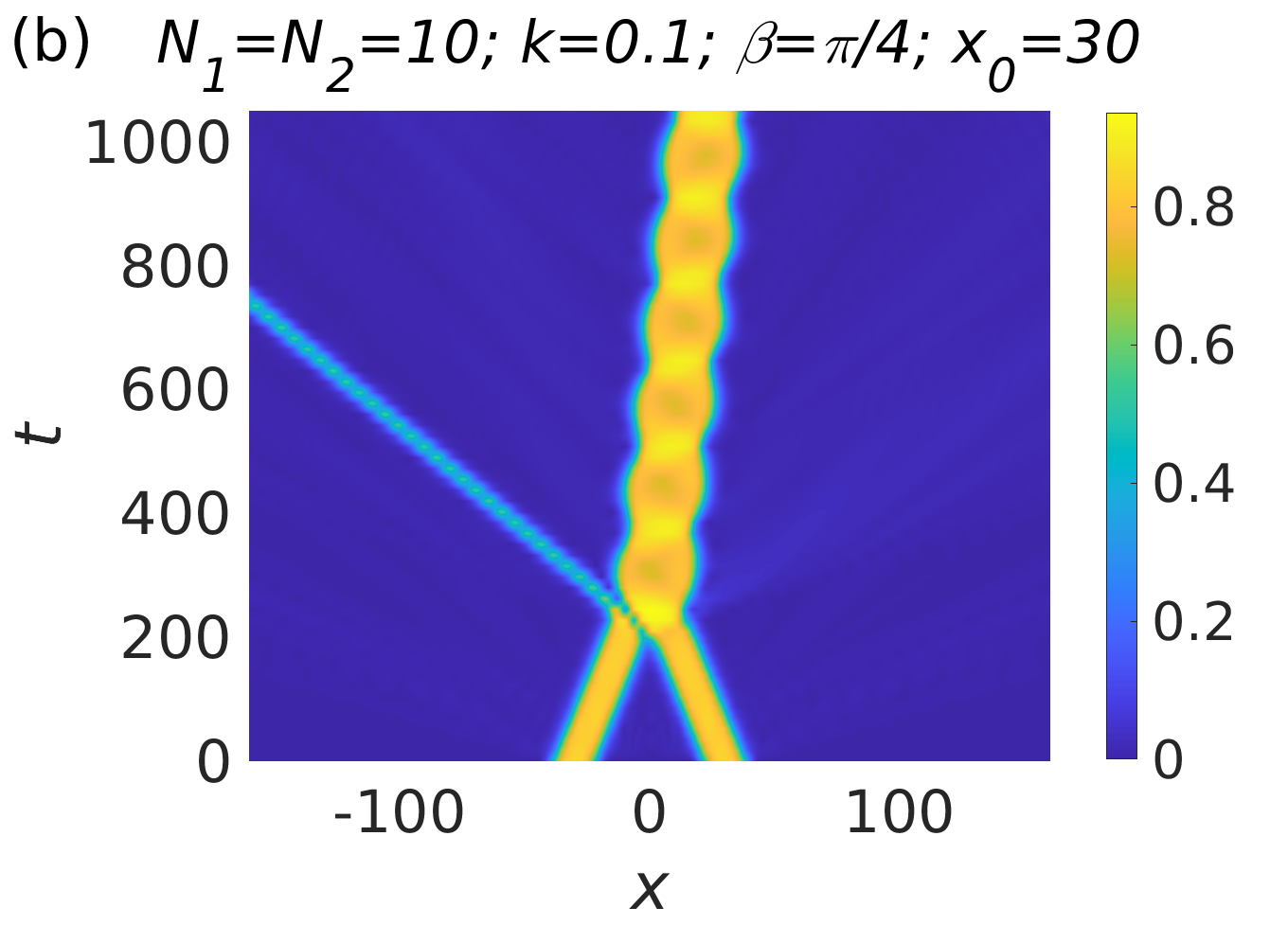}}
    \centerline{\includegraphics[width=4.55cm]{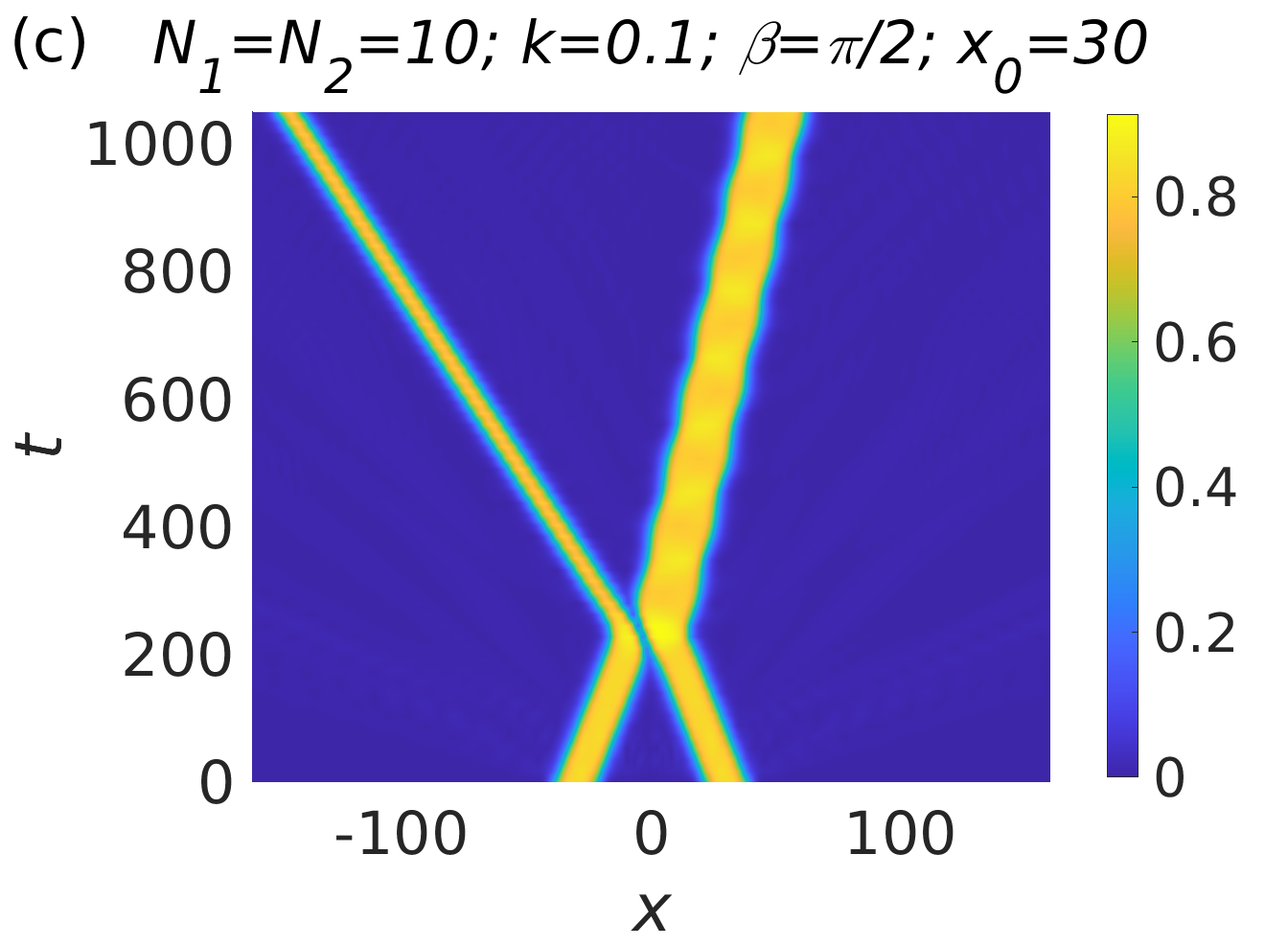} \includegraphics[width=4.55cm]{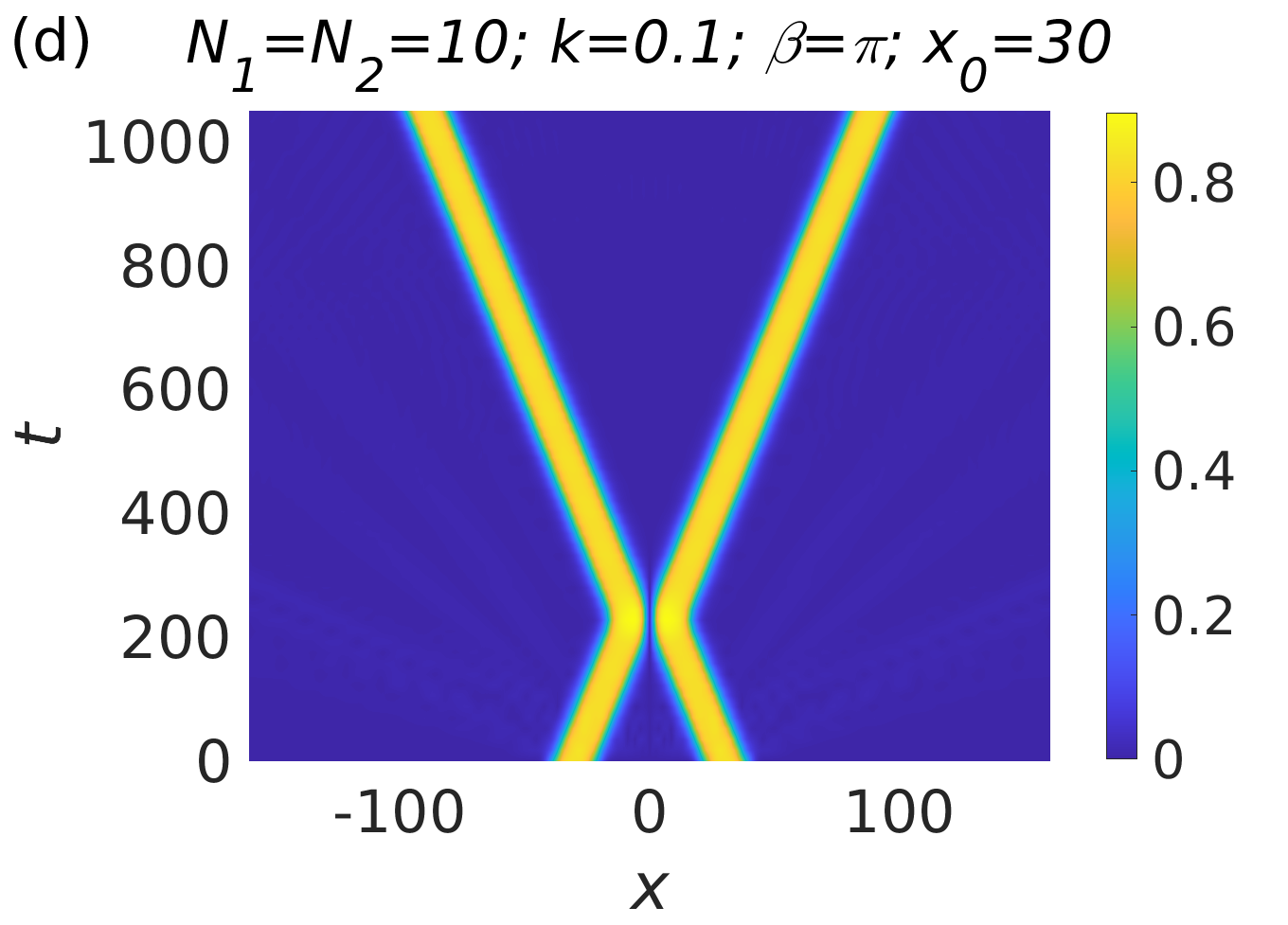}}
\caption{The same as in Fig.~\ref{fig-9abcd}, but this figure presents the scenario of the interaction of slowly moving flat-top QDs with an initial momentum of $k=0.1$ and a starting position at $x_0=30$.}
\label{fig-11abcd}
\end{figure}

Figure~\ref{fig-11abcd} displays the scattering dynamics of two identical large flat-top QDs moving slowly in opposite directions. In the context of slow motion, the behaviour of these large in-phase flat-top QDs as shown in Fig.~\ref{fig-11abcd}(a) closely resembles the interaction observed in non-moving droplets, as depicted in Fig.~\ref{fig-9abcd}(a). During the slow-moving scenario, the main part of the atoms remains confined within the newly created central droplet, with only relatively minor amounts of particles being emitted as small amplitude waves. The combined central QD exhibits excitation, manifesting in amplitude oscillations. 
%

A symmetry breaking occurs in the relative phase difference values within the interval $0 < \beta \leq \pi/2$. The collision dynamics of quantum droplets for given values of $\beta$, namely $\pi/4$ and $\pi/2$, are depicted in Figs.~\ref{fig-11abcd}(b) and (c), respectively. This behaviour closely aligns with the dynamics observed in the interaction of two stationary quantum droplets, as shown in Fig.~\ref{fig-9abcd}(b). 
In these instances, incoming QDs do not maintain their shape after the collision. 
Figure~\ref{fig-11abcd}(d) shows the collision of two large flat-top droplets characterized by an out-of-phase structure $\beta=\pi$. This collision reveals that when large QDs with an equal normalization meet, they rebound instantaneously, creating a final trajectory resembling a complete mirror reflection. 

\begin{figure}[htbp]
  \centerline{ \includegraphics[width=4.55cm]{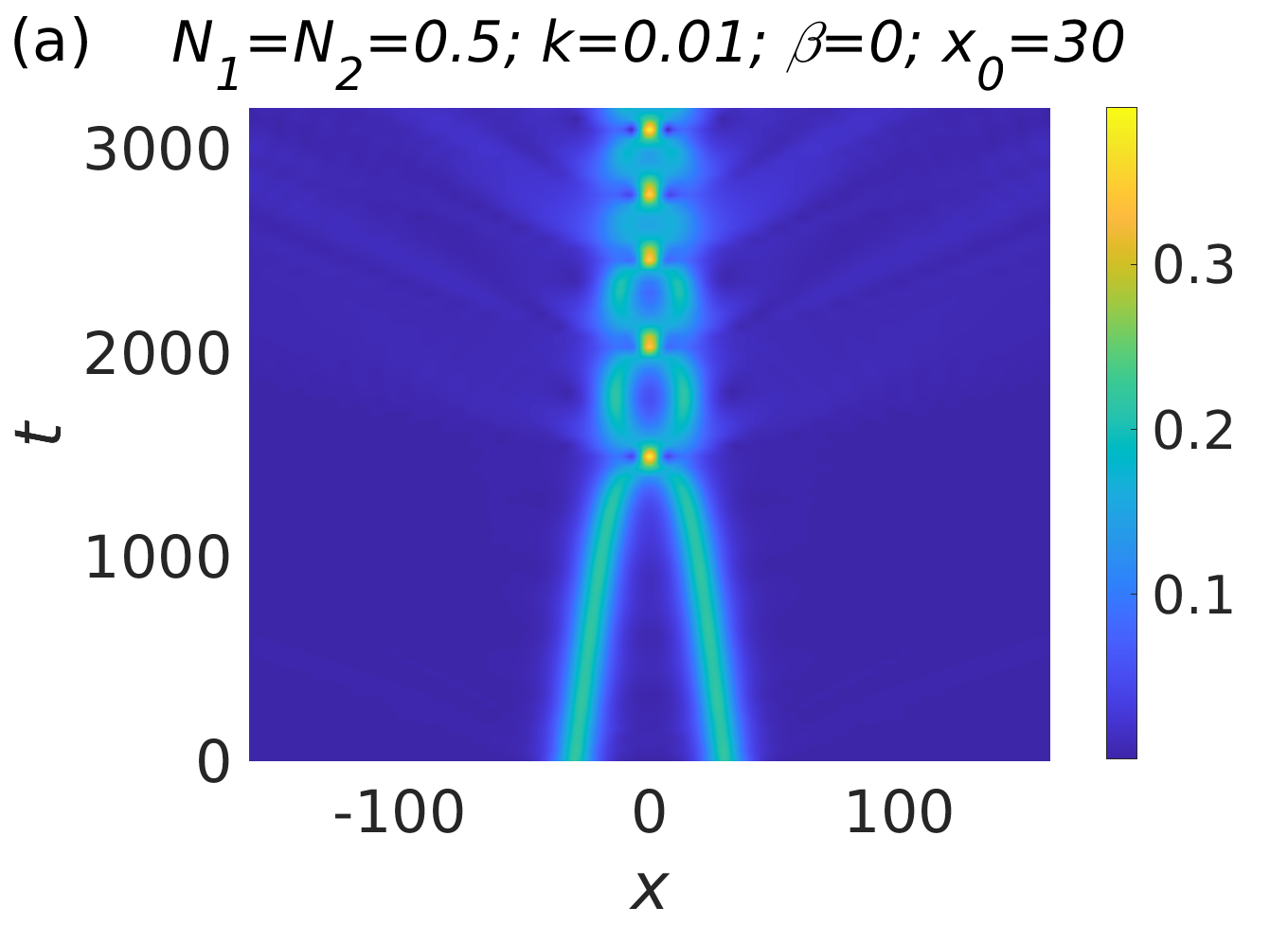} \includegraphics[width=4.55cm]{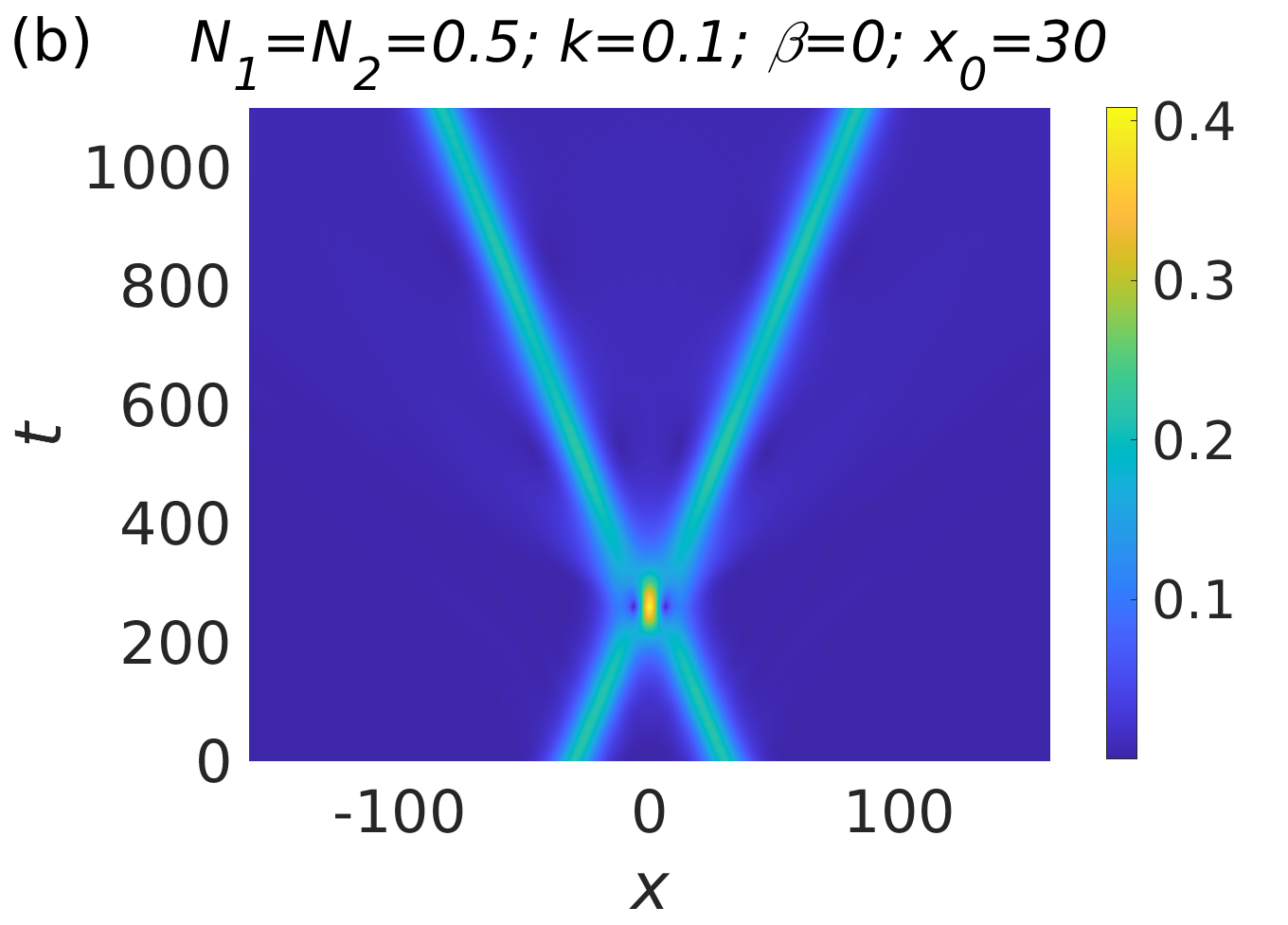}}
  \centerline{ \includegraphics[width=4.55cm]{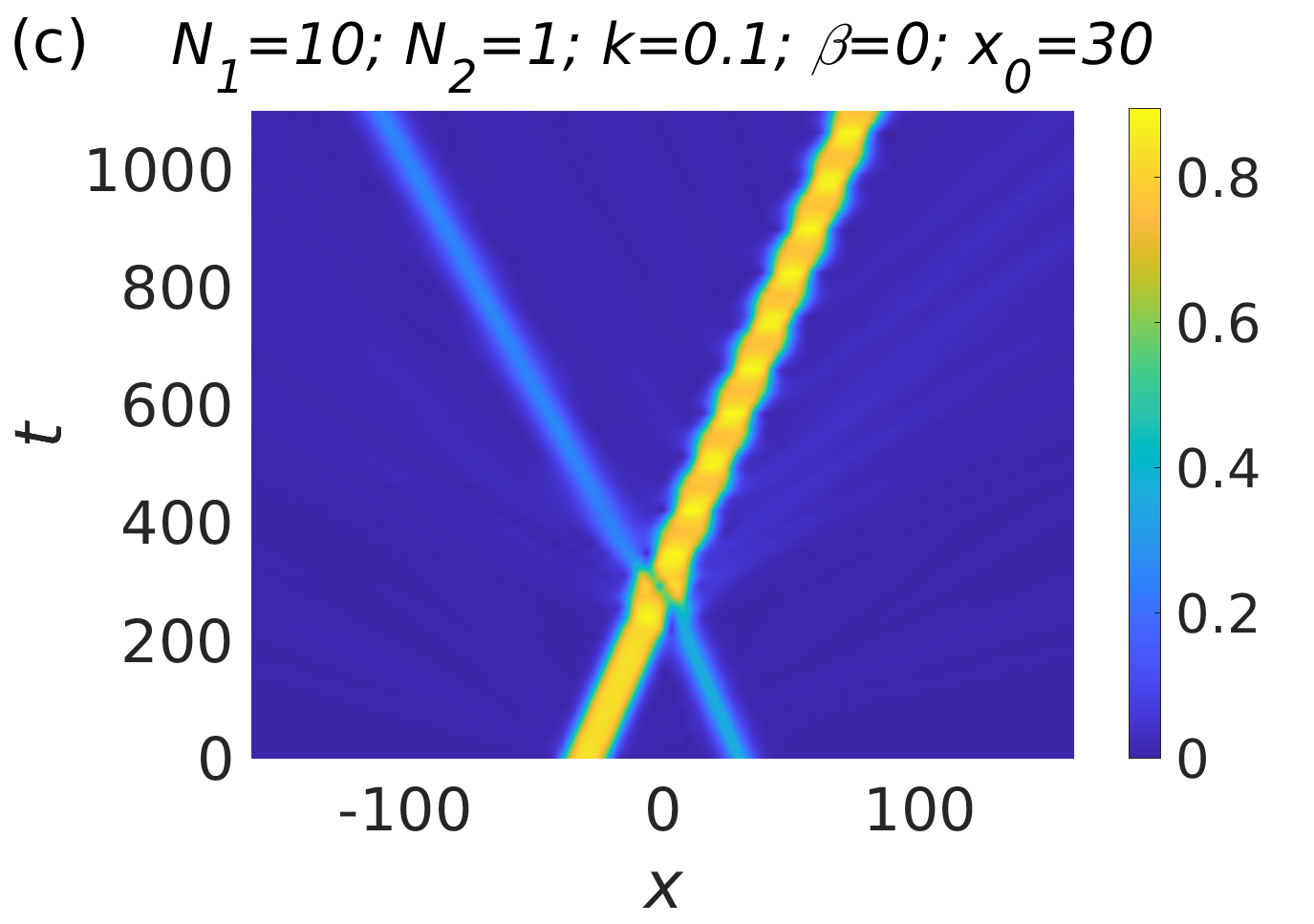} \includegraphics[width=4.55cm]{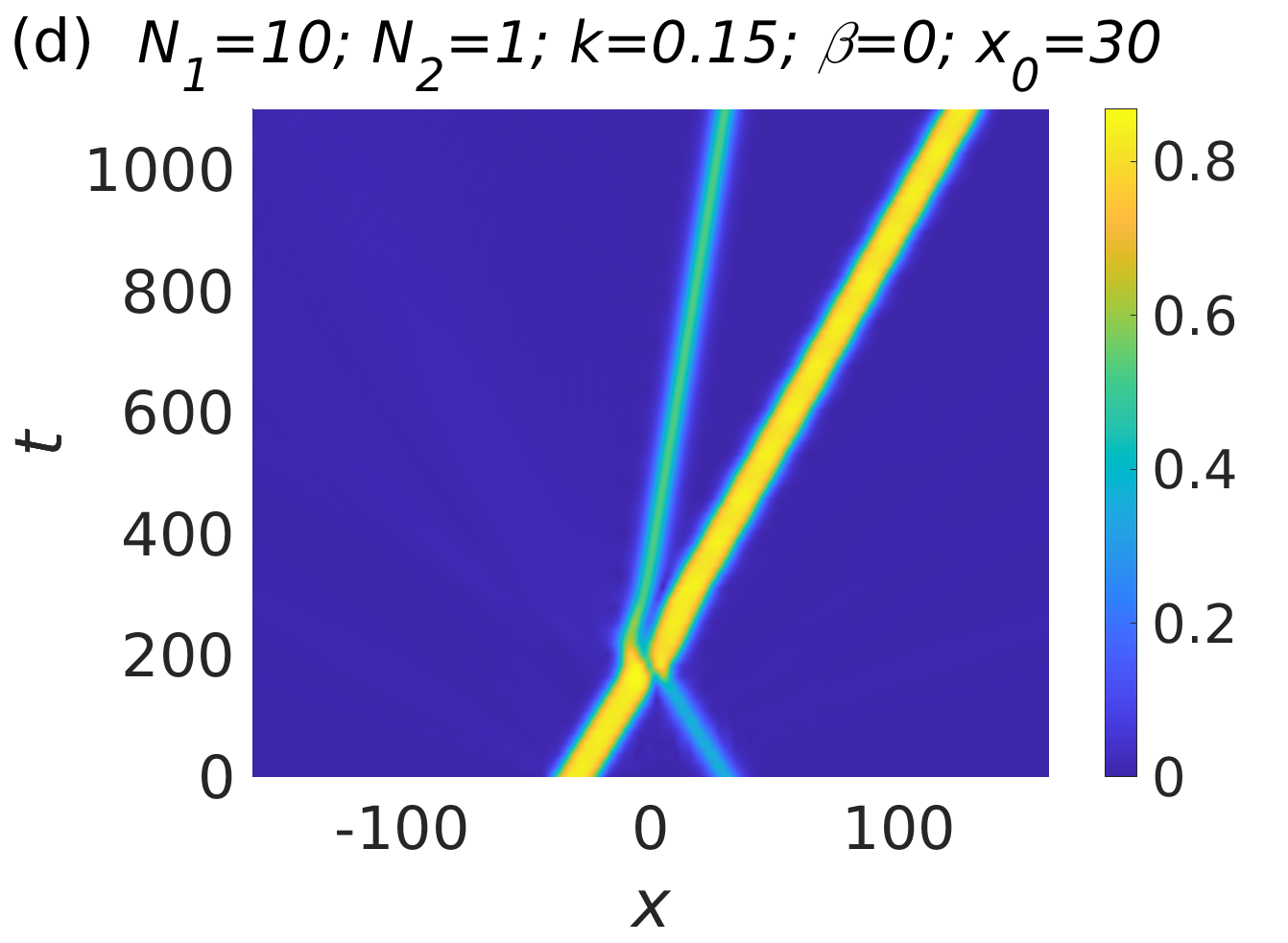}}
\caption{Time-dependent density profiles of two in-phase QDs are presented in this sequence. Panel (a) shows the case for $N_1=N_2=0.5$ and momentum $k=0.01$ less than the critical value $k_{cr}$, while panel (b) displays a similar setup as in panel (a) but with $k=0.1$ exceeding the critical value $k_{cr}$. Panel (c) illustrates the collision dynamics of QDs with differing norms, $N_1=10$ and $N_2=1$, at a momentum of $k=0.1$. Panel (d) follows the same parameters as in panel (c) but with a momentum of $k=0.15$.}
\label{fig-12abcd}
\end{figure}

We now examine the collision dynamics between two identical bell-shaped quantum droplets with a small number of particles at low speeds. When the relative phase difference $\beta$ equals zero, the collision dynamics of these QDs are influenced by their initial velocities. If the initial momentum of the droplets falls below a critical threshold $k<k_{cr}$, they merge. However, this merging does not result in the complete fusion of the colliding droplets into a single entity. Instead, it leads to the formation of a bound state, referred to as a weak merging regime, as illustrated in Fig.~\ref{fig-12abcd} (a). Following the initial collision, the droplets reemerge as two distinct localized waves subsequently undergoing multiple collisions. This type of behaviour is also present in various other nonlinear systems. For example, similar investigations have been reported in the context of 1D cubic-quintic media in Ref.~\cite{Khaykovich2006}.

When the initial momentum of the droplets exceeds a critical threshold $k>k_{cr}$, each individual droplet remains intact after the collision, as depicted in Fig.~\ref{fig-12abcd} (b). For other different values of $\beta$, the collision behaviour of small bell-shaped droplets at low speeds closely resembles the dynamics observed in large flat-top QDs, as shown in panels (b), (c) and (d) of Figs.~ \ref{fig-11abcd}. 

In addition to our analysis of colliding droplets with equal norms, we also investigate collisions involving droplets with differing particle numbers. Panels (c) and (d) in Fig.~\ref{fig-12abcd} depict the collision between a large droplet with N = 10 and a smaller one with N = 1. The subsequent trajectory of the smaller droplet after the collision highly depends on its initial velocity. In Fig.~\ref{fig-12abcd}(c) and (d), we explore a scenario where large flat-top and small Gaussian-like droplets collide with an initial momentum of $k=0.1$. Following the collision, either one or both of the droplets may become excited and disperse, displaying internal periodic oscillations. When the initial speed varies, the droplets may separate after the collision, potentially resulting in a redistribution of particle counts. Remarkably, at specific initial momentum, such as $k=0.3$, a droplet with a limited number of particles changes its direction, moving in the opposite direction towards a larger QD as shown in Fig.~\ref{fig-12abcd}(d).
Conservation of norm and momentum are monitored by calculation of these quantities before and after a collision. For norm $N$ we use Eq.~(\ref{norm}) and for momentum $M$ the following expression~\cite{Barenghi2016}
\begin{equation}
   M \equiv i\int_{-\infty}^{\infty} (\Psi^* \Psi_x -\Psi \Psi^*_x) dx.
\label{momentum}
\end{equation}
The values were calculated at $t =0$, (initial moment of time)  and $t =360$ (after collision), and they were the same $N=11$, and $M= 5.34$ for both mentioned moments of time.

\section{Conclusions}
\label{sec:conc}
	In conclusion, this paper presents a systematic theoretical study of quantum droplets in quasi-one-dimensional Bose-Einstein condensates comprising a symmetric mixture of dilute cold atomic gases. We introduce a modified mean-field one-dimensional Gross-Pitaevskii equation with cubic and quartic nonlinearity terms, accounting for attractive two-body interactions and the repulsive impact of quantum fluctuations. Dynamical variational equations for time-dependent parameters of quantum droplets are derived using a super-Gaussian ansatz. Based on these equations, we obtain approximate stationary solutions for quantum droplets, which exhibit a bell-shaped solitonic form for smaller norms. In comparison, profiles for larger norms become wider, with amplitude and chemical potential reaching the Thomas-Fermi limit. The time-dependent variational equations are explored to study the stability of quantum droplets and oscillations of their shape near stationary solutions. The frequencies of width and amplitude oscillations are calculated, demonstrating that oscillations of width and shape are in counter-phase, ensuring norm conservation. It is also revealed that droplets are stable to small variations in shape. The approximate results for stationary solutions and their stability, obtained by applying variational equations, are in excellent agreement with the results of direct numerical simulations of Eq.~(\ref{quasi_1D_gpe}).

The dynamics of quantum droplets at periodic modulation of the atomic scattering length are investigated. The phenomenon is described by periodic modulations of the nonlinear coefficients $\gamma(t)$ and $\delta(t)$.  The analysis is performed using variational equations, demonstrating that relatively small modulation amplitudes can initiate forced oscillations and resonances. The variational approximation accurately predicts the parameters of oscillations and resonance frequencies. For larger values of modulation amplitudes, direct numerical simulations of the modified GPE show that quantum droplets emit small density waves, eventually decaying and disappearing.

	The particular case of the LHY fluid is discussed using the variational method with a super-Gaussian ansatz, assuming that in the modified GPE, the repulsive quartic term induced by quantum fluctuations is dominant, and the mean-field cubic nonlinearity term is neglected. With the LHY term being repulsive and balanced by a parabolic trap, it is shown that stable stationary quantum LHY fluid exists. 

	The last part of the paper is devoted to the interaction and collision of two quantum droplets through direct numerical solutions of the modified quasi-1D GPE. It concludes that the interaction between two quantum droplets strongly depends on their distance, relative phase, initial velocities, and norms. Two in-phase identical droplets with zero initial velocity attract each other if their initial locations are close enough to interact by tails. Repulsion occurs when the relative phase is equal to $\pi$. For other phase differences, interaction may be asymmetric, with redistribution of density, merging, or separation of droplets after collision. In summary, the interaction of quantum droplets exhibits complex, non-integrable behaviour.

\section*{Acknowledgements}
This work has been funded from the state budget of the Republic of Uzbekistan.


\end{document}